\documentclass{article}

\usepackage{latexsym}
\usepackage{amsmath}
\usepackage{amsfonts}
\usepackage{theorem}
\usepackage{epsfig}

\newtheorem{theorem}{Theorem}
\newtheorem{lemma}[theorem]{Lemma}

\newtheorem{proposition}[theorem]{Proposition}
\newtheorem{corollary}[theorem]{Corollary}

\newtheorem{remark}[theorem]{Remark}

\newcommand{\zerarcounters} { \setcounter{equation}{0}
\setcounter{theorem}{0} }

\newcommand{\und}{\underline}

 \newcommand{\undm}{\underline{m}}
\newcommand{\undn}{\underline{n}}

\newcommand{\undomega}{\underline{\omega}}

\newcommand{\calA}{{\mathcal A}} \newcommand{\calB}{{\mathcal B}}

\newcommand{\calI}{{\mathcal I}} 
\newcommand{\calK}{{\mathcal K}}

\newcommand{\calQ}{{\mathcal Q}} \newcommand{\calR}{{\mathcal R}}
\newcommand{\calS}{{\mathcal S}} \newcommand{\calT}{{\mathcal T}}
\newcommand{\calU}{{\mathcal U}}

\newcommand{\bc}{\begin{center}} \newcommand{\ec}{\end{center}}

\newcommand{\eps}{\epsilon}
\newcommand{\EndofStatement}{\samepage\hfill\Box}
 \newcommand{\QED}{\Fullbox}

\newcommand{\novomod}[1]{\ll\!\! #1 \!\!\gg}

\newcommand{\Fullbox}{\hfill{\rule{1.8mm}{1.8mm}}}
\newcommand{\til}{\tilde} 

\newcommand{\defi}{\; := \; }
\newcommand{\infinito}{\infty}
\newcommand{\q}[2]{\left( #1 \, \left| \, #2 \right. \right)}%
\newcommand{\qq}[3]{\left(#1\,\left|\,#2\,\left|\,#3\right.\right.\right)}
\newcommand{\qqq}[4]{\left(#1\,\left|\,#2\,\left|\,#3\,\left|\,#4\right.\right.\right.\right)}
\newcommand{\Proof}{\vspace{-0.0cm} \noindent {\bf Proof. }\ }%

\newcommand{\Z}{\mathbb Z} \newcommand{\N}{\mathbb N}
\newcommand{\R}{\mathbb R} \newcommand{\C}{\mathbb C}

\newcommand{\ren}[1]{\,{\mathfrak R}_{#1}}%

\renewcommand{\bar}{\overline}%

\begin{document}

\begin{flushleft}
{\noindent \bf \Large
  Perturbative Analysis of Dynamical Localisation
} 
\end{flushleft}

\vspace{0.1cm}

\begin{center}

J. C. A. Barata\footnote{Partially supported by CNPq. E-mail:
  jbarata@if.usp.br}
and
D. A. Cortez\footnote{Work supported by FAPESP. E-mail:
  dacortez@fma.if.usp.br}

\end{center}

\begin{quote}
{\small
\noindent{\bf Abstract.} In this paper we extend previous results on
convergent perturbative solutions of the Schr\"odinger equation of a
class of periodically time-dependent two-level systems. The situation
treated here is particularly suited for the investigation of two-level
systems exhibiting the phenomenon of (approximate) dynamical
localisation. We also present a convergent perturbative expansion for
the secular frequency and discuss in detail the particular case of
monochromatic interactions (ac-dc fields), providing a complete
perturbative solution for that case. Our method is based on a
``renormalisation'' procedure, which we develop in a more systematic
way here. For being free of secular terms and uniformly convergent in
time, our expansions allow a rigorous study of the long-time behaviour
of such systems and are also well-suited for numerical computations,
as we briefly discuss, leading to very accurate calculations of
quantities like transition probabilities for very long times compared
to the cycles of the external field.  }
\end{quote}


\section{General Description and Previous Results}
\label{sec:intro} 
\zerarcounters

The study of periodically or quasi-periodically time-dependent
two-level systems is of basic importance for many physical
applications, ranging from condensed matter physics to quantum optics,
as in problems of the theory of spin resonance, in problems of quantum
tunnelling or in the semi-classical theory of the laser.  They can be
used, for instance, to describe the behaviour of a spin $1/2$ system
in a time-dependent magnetic field, in which case the corresponding
Schr\"odinger equation takes the form (we adopt $\hbar=1$)
\begin{equation}
i\partial_t\Psi \;= \;H(t)\Psi\,, 
\qquad \mbox{ with } \qquad 
H(t) \; = \; -\frac{1}{2}\vec{B}(t) \cdot \vec{\sigma} \; ,
\label{EquacaodeSchroedinger}
\end{equation}
where $\Psi(t) = \left( { \psi_1 (t) \atop \psi_2 (t)} \right) \in
\C^2 $, $\vec{B}(t)= (B_1(t), \,B_2(t), \, B_3(t))$ and $\vec{\sigma}
= (\sigma_1 , \, \sigma_2 , \, \sigma_3)$ are the Pauli matrices.

Systems like this have been analysed by many authors in various
approximations, as in the pioneering works of Rabi \cite{Rabi}, of
Bloch and Siegert~\cite{BlochSiegert} and of Autler and
Townes~\cite{AutlerTownes} (see
also~\cite{Walter1,WreszinskiCasmeridis} for more recent discussions).
We should remark, however, that the interest in the solutions of
(\ref{EquacaodeSchroedinger}) is not restricted to the investigation
of quantum systems.  As first pointed by Feynman, Vernon and
Hellwarth~\cite{FeynmanVernonHellwarth} (see also the recent
discussion in~\cite{BagrovBarataGitmanWreszinski}), the quantum system
(\ref{EquacaodeSchroedinger}) is equivalent to the
\underline{classical} Hamiltonian system describing a classical
gyromagnet precessing in a magnetic field: $ \frac{d}{dt}\vec{\calS} =
- \vec{B}(t) \times \vec{\calS} $, where $\vec{\calS}$ is a unit
vector.

Of particular interest is the situation where the Schr\"odinger
equation takes the form
\begin{equation}
i\partial_t \Psi(t) \; = \; H_1(t)\Psi(t) , \qquad \mbox{ with }
\qquad H_1(t) \; := \; \eps \sigma_3 - f(t)\sigma_1 ,
\label{equacaodeSchroedingerparaPsi}
\end{equation}
where $f(t)$ is a function of time $t$ and $ \eps\in\R$ is
constant. By a time-independent unitary transformation,
representing a rotation of $\pi/2$ around the 2-axis, we get the
equivalent system
\begin{equation}
 i\partial_t \Phi(t) \; = \; H_2(t)\Phi(t),
\qquad \mbox{ with } \qquad H_2 (t) \; := \; \eps \sigma_1 +
f(t)\sigma_3 , \label{equacaodeSchroedingerparaPhi}
\end{equation}
where $ \Phi (t) := \exp\left(-i\pi\sigma_2 /4\right)\Psi(t) $
and $H_2 (t) := \exp\left(-i\pi\sigma_2 /4\right) \, H_1(t)\,
\exp\left(i\pi\sigma_2 /4\right)$.

One can either interpret the system
(\ref{equacaodeSchroedingerparaPsi}) as describing a spin $1/2$ system
as (\ref{EquacaodeSchroedinger}) under a magnetic field $\vec{B} =
(2f(t), \, 0, \, -2\eps)$, or as a system with an unperturbed diagonal
Hamiltonian $H_0 := \eps \sigma_3 $, representing a two-level system
with energy levels $\pm\eps$, subjected to a time-dependent
perturbation $H_I(t) := - f(t)\sigma_1$, inducing a time-depending
transition between the unperturbed eigenstates of $H_0$. The
equivalent system (\ref{equacaodeSchroedingerparaPhi}), in turn,
represents either a spin $1/2$ system as (\ref{EquacaodeSchroedinger})
under a magnetic field $\vec{B} = (-2\eps, \, 0, \, -2f(t))$, or a
two-level system composed by two uncoupled (for $\eps=0$) orthogonal
time-dependent states $\exp\left(-i\int_0^t f(\tau)d\tau\right)
{\left(1 \atop 0\right)}$ and $\exp\left(+i\int_0^t
  f(\tau)d\tau\right) {\left(0 \atop 1\right)}$, subjected to a
constant perturbation $\eps \sigma_1$ inducing a transition between
them.

To explain the purpose of the present paper, we have to describe some
of our previous results.  In~\cite{qp} and~\cite{pp} we studied the
system described by (\ref{equacaodeSchroedingerparaPsi}) or
(\ref{equacaodeSchroedingerparaPhi}) in the situation where $f$ is a
periodic or quasi-periodic function of time and $\eps$ is ``small''.
It is well know that the usual perturbative approach, based, f.i., on
the Dyson series, leads to difficulties involving secular terms (i.e.,
polynomials in $t$ that appear order by order in perturbation theory
and spoil the uniform convergence (in $t$) of the perturbative series)
and, for quasi-periodic interactions, small denominators. This last
problem is typical of perturbative approximations for solutions of
differential equations with quasi-periodic coefficients and is
well-known as one of the main sources of problems in the
mathematically precise treatment of such equations.

In~\cite{qp} and~\cite{pp}, a special perturbative expansion (power
series expansion in $\eps$) was developed, whose main virtue is to be
free of secular terms. The algorithm employed involves an inductive
``renormalization'' of a sort of effective field introduced through an
exponential Ansatz (the function $g$, to be introduced below).  For
the sake of the reader we will shortly recall our method of
elimination of secular terms in Section~\ref{sec:elimination}.  In the
general case where $f$ is {\it quasi-periodic}, it was established in
\cite{qp} that the coefficients of the expansion are also
well-defined~quasi-periodic functions of time but, due mainly to the
presence of small denominators, we were not able to prove convergence
of our $\eps$-expansion. Actually, a convergent power expansion in
$\eps$ is not expected without further assumptions (for a detailed
analysis of these issues in related systems,
see~\cite{GallavottiGentile}).

Less problematic is the situation where $f$ is a {\it periodic}
function, when the obstacle represented by the small denominators
is naturally absent. In~\cite{pp}, we showed how the difficulties
analysed in~\cite{qp} can be circumvented in the case of
\underline{periodic} $f$ and we were able to establish the
convergence of our perturbative $\eps$-expansion uniformly in
$t\in \R$.

As discussed in \cite{pp}, our method not only recovers the Floquet
form of the solution of the time-depending Schr\"odinger equation (see
(\ref{U11U12})-(\ref{u11u12}) below), but also allows the computation
of the secular frequency and of the Fourier coefficients in terms of
explicit {\it convergent} $ \eps$-expansions, what constitutes a
feature of our algorithm, compared to other expansion methods.

Due to the technical difficulties involved, we restricted our analysis
in~\cite{pp} to two classes of periodic functions, namely those
satisfying the conditions (I) or (II) presented below (see~\cite{pp}).
Our purpose in the present paper is to extend the results of~\cite{pp}
to an additional class of periodic functions.  The inclusion of this
additional class leads to a essentially complete perturbative solution
for some simple periodic functions, as $f(t) = F_0 +\varphi 
\cos(\omega t)$, representing the important case of a monochromatic 
interaction (also known as ac-dc fields).  

The situation we treat here is also relevant for the rigorous
discussion of the phenomenon of dynamical localisation, also known
(less properly) as coherent destruction of tunnelling.  This
phenomenon, first pointed in~\cite{Grossmann}, indicates the
possibility to (approximately) freeze the initial state of a quantum
system through the action of a suitable external time-dependent
interaction. This effect has been the object of various recent
investigations. In~\cite{Sacchetti}, for instance, a rigorous general
criterion for the occurrence of dynamical localisation was established
and applied to interesting situations, like the ac-dc field and the
bichromatic field. Some of the conclusions of~\cite{Sacchetti} on the
ac-dc field are indirectly reproduced in Section~\ref{sec:example},
below.  We refer the reader to~\cite{BarataWreszinski2}
and~\cite{Sacchetti} for more references on this subject.

The main result of~\cite{pp} can be captured in the next theorem,
for whose statement we need a definition we will repeatedly
use in this work: for an almost periodic function $ h$ we denote
by $M(h)$ the ``mean value'' of $h$, defined as
\begin{equation} \label{eq:media}
M(h) \; := \; \lim_{T\to\infty} \frac{1}{2T}\int_{-T}^{T} \, h(t) \, dt
\, .
\end{equation}
We remark that the limit in
(\ref{eq:media}) is always well defined for any quasi-periodic
function $h$.  The mean value $ M(h)$ equals the constant term in the
Fourier expansion of $h$.  Details can be found
in~\cite{Katznelson,Corduneanu}.

\begin{theorem} \label{teo:floquet}
Let $ f$ be a real $T_\omega$-periodic function of time ($T_\omega
:= 2\pi/\omega$ with $\omega >0$) whose Fourier decomposition
$f(t) = \sum_{ n \in \Z } F_{n} e^{i n \omega t}$,
contains only a finite number of terms, i.e., the set of integers
$\{ n\in\Z | \; F_n \neq 0\}$ is a finite set. 
Let
\begin{equation} \label{solucaodaequacaodeSchroedinger}
\Phi(t) = 
  \left(
    \begin{array}{c}
       \phi_+ (t) \\ \phi_- (t)
    \end{array}
  \right) = U(t)\Phi(0) =  U(t, \, 0)\Phi(0)
\end{equation}
be the solution of the Schr\"odinger equation
(\ref{equacaodeSchroedingerparaPhi}).
Consider the two following distinct conditions on $f$:
\begin{enumerate}
\item[\rm (I)] $M(\calQ_0)\neq 0$.

\item[\rm (II)] $M(\calQ_0) = 0$ but $M(\calQ_1) \neq 0$, where
\begin{equation}
       q(t) := \exp\left( i\int_0^t \, f(\tau)d\tau \right), 
\;\;
\calQ_0(t)  :=  q(t)^2 \; = \; \exp\left( 2i\int_0^t \, f(\tau)d\tau
\right) 
\label{definicaodecalQ0}
\end{equation}
and
\begin{equation}
  \calQ_1(t) \; := \; \calQ_0(t) \int_0^t \left(
  \calQ_0(\tau)^{-1}-M\left(\calQ_0^{-1}\right)\right) d\tau .
\label{definicaodecalQ1}
\end{equation}
\end{enumerate}
Then, for each $ f$ as above, satisfying condition (I) or (II),
there exists a constant $K >0$ (depending on the Fourier
coefficients $\{ F_n , \; n\in \Z \; , n\neq 0\}$ and on $\omega$)
so that, for each $ \eps$ with $|\eps| < K$, there are $\Omega
\in \R$ and $T_\omega$-periodic functions $u_{11}^\pm$ and
$u_{12}^\pm$ such that the propagator $ U(t)$ of
(\ref{solucaodaequacaodeSchroedinger}) can be written as
\begin{equation*}
U(t) \; = \; \left(
 \begin{array}{cc}
    U_{11}(t) & U_{12}(t) \\  & \\ U_{21}(t) & U_{22}(t)
   \end{array}
\right) \; = \; \left(
 \begin{array}{cc}
    U_{11}(t) & U_{12}(t) \\ & \\ -\overline{U_{12}(t)} & \overline{U_{11}(t)}
   \end{array}
\right) , 
\end{equation*}
with
\begin{equation}
U_{11}(t) \, = \, e^{-i\Omega t}\, u_{11}^-(t) + e^{i\Omega t}\,
u_{11}^+ (t), \quad U_{12}(t) \, = \, e^{-i\Omega t}\, u_{12}^-
(t) + e^{i\Omega t}\, u_{12}^+ (t). \label{U11U12}
\end{equation}
The functions  $u_{11}^\pm$ and $u_{12}^\pm$ have absolutely and
uniformly converging Fourier expansions
\begin{equation}
u_{11}^\pm(t) \; = \; \sum_{n\in \Z} \calU_{11}^{\pm}(n)e^{i n
\omega t}, \qquad u_{12}^\pm(t) \; = \; \sum_{n\in \Z}
\calU_{12}^{\pm}(n)e^{i n \omega t}. \label{u11u12}
\end{equation}
Moreover, under the same assumptions, $ \Omega$ and the Fourier
coefficients $ \calU_{11}^{\pm}(n)$ and $ \calU_{12}^{\pm}(n)$ can
be expressed in terms of absolutely converging power series on $
\eps$. $\EndofStatement\!\!$ \label{Teoremaprincipalsobreopropagador}
\end{theorem}


Let us now discuss the conditions (I) and (II) of Theorem
\ref{Teoremaprincipalsobreopropagador}.  Writing the Fourier
decomposition of $ f$ as $ f(t) = F_0 + \sum_{n=1}^{J} \left[
  \varphi_1^{(n)} \cos( n \omega t) + \varphi_2^{(n)} \sin( n \omega
  t) \right] $ the set ${\mathfrak F}_{J, \, \omega} $ of all possible
functions $f$ with a given $ J$ and $ \omega$ can be identified with
the parameter space $ \R^{2J+1}$ of all real coefficients $F_0, \;
\varphi_{1, \, 2}^{(n)}$, $1\leq n \leq J$.
The (complex) condition $ M(\calQ_0)=0$ determines a $(2J)$ or
$(2J-1)$-dimensional subset of $ {\mathfrak F}_{J, \, \omega}$,
where condition (II) eventually applies. It is also on this subset
that the more restrictive condition $ M(\calQ_0) = M(\calQ_1)=0$
should hold, restricting the parameter space of $ f$ to a
$(2J-1)$, $(2J-2) $ or $(2J-3)$-dimensional subset, if it is
non-trivial.  One should, therefore, expect that successive
conditions like (I) and (II) would eventually exhaust completely
the set $ {\mathfrak F}_{J, \, \omega}$.

To illustrate all this, let us consider the simplest example, when $
f$ represents a monochromatic interaction: $ f(t) = \varphi_1
\cos(\omega t) + \varphi_2 \sin(\omega t)$, with $(\varphi_1, \,
\varphi_2) \in \R^2$. A simple computation shows that $ M(\calQ_0) =
e^{2i\gamma_f}J_0 \left(\frac{2\varphi_0}{\omega}\right), $ where
$\varphi_0 := \sqrt{\varphi_1^2 + \varphi_2^2}$, $ J_0$ is the Bessel
function of first kind and order zero and, in this case, $ \gamma_f =
\varphi_2/\omega $. Moreover, $M(\calQ_1)=0$ for \underline{all}
$(\varphi_1, \, \varphi_2) \in \R^2$. (See \cite{pp} for details).
Hence, condition (I) is satisfied for all $(\varphi_1, \, \varphi_2)
\in \R^2$, except in the circles defined by $\varphi_0=\omega x_a/2$,
$ a = 1, \,2, \ldots$, where $ x_a$ if the $a$-th zero of $ J_0$ in $
\R_+$. Condition (II), however, is never fulfilled in this case. To
achieve a complete solution we have, therefore, to extend Theorem
\ref{Teoremaprincipalsobreopropagador} to include further conditions
beyond (I) and (II), holding on the circles $\varphi_0=\omega x_a/2$.

The purposes of this paper are to identify the first condition
following (I) and (II), which we call condition (III), to show that
the method of elimination of secular terms holds in this case as well
and, for periodic interactions, to show that the expansion
(\ref{eq:ansatz}) converges for $|\eps|$ sufficiently small, uniformly
for $t\in\R$. As we will discuss, this leads to a complete
perturbative solution for the monochromatic interaction.  As we will
see, the identification of condition (III) and the application of the
method of elimination of secular terms to it are highly non-trivial
tasks.

\begin{center}
*
\end{center}

This paper is organised as follows. 
In Section~\ref{sec:elimination} we describe the general strategy
employed to eliminate the secular terms and present our main theorems.
In Section~\ref{sec:PMVRP} we prove some interesting and useful
mathematical results on the mean value of some quasi-periodic
functions, introduce the ``renormalisation'' operation, important to
organise our procedures, and introduce some notations we will use
throughout the paper.
In Section~\ref{sec:case}, which is the technically central piece or
this work, we apply our strategy of elimination of the secular terms
to the situation we wish to analyse.
In Section~\ref{sec:example} we apply our results to case of
monochromatic interactions and discuss the issue of dynamical
localisation in this case.
In Section~\ref{sec:numerical} we briefly describe some numerical
calculations based on our results.  Sections \ref{sec:example} and
\ref{sec:numerical} contain some of the physical applications of our
work.
In Appendix~\ref{ap:VariasProvas} we treat several results used
in Section~\ref{sec:case}, whose proofs unfortunately involve too many
algebraic computations to be included in the main text.
Appendix~\ref{sec:convergence} is dedicated to the proof of convergence
of our expansions for the periodic case. 
Appendix~\ref{app:novo} contains some comments on the Fourier
coefficients of the wave functions.
Finally, Appendix~\ref{identidadeBessel} sketches the proof of an
identity on Bessel functions we use in  Section~\ref{sec:example}.


\subsection{Comments on the Notation}
\label{ssec:notation}

In this paper, $\Z_+$ will denote the set of all non-negative integers
(zero included) and $\Z_\ast$ the set of all integers, excluding zero.
$\Z_{\ast+}$ is the set of all positive integers. These notations
are also applied to $\Z^A$ and to the real line $\R$.

Vectors in $\Z^A$ (or $\R^A$) will be written as $\und{v}$. The
operation $\und{v} \cdot \und {u}$ will denote the scalar product
in $\Z^A$ (or $\R^A$), defined as $\und{v} \cdot \und{u} \defi v_1
u_1 + \cdots + v_A u_A$.

For a quasi-periodic function $h: \R \to \C$, we write its Fourier
decomposition as
$
h(t) = \sum_{\und{m} \in \Z^A} H_{\und{m}} e^{i \und{m}
\cdot \und{\omega}_h t} \, ,
$
where $A$ is some positive integer and $\und{\omega}_h \in \R^A_+$ 
(see, e.g.,~\cite{Katznelson, Corduneanu}). The
Fourier coefficient $H_{\und{0}}$ will be denoted simply as $H_0$.

For $ m\in \Z$ we denote by $ \novomod{m}$ the following function:
\begin{equation}
\novomod{m} \; := \;  
\left\{ 
  \begin{array}{cl}
                    |m|, & \mbox{for } m \neq
                    0\\ 1 ,& \mbox{for } m =0 
  \end{array}
\right. .
\label{definicaodenovomod}
\end{equation}

Beyond the functions $\calQ_0$ and $\calQ_1$ defined in 
(\ref{definicaodecalQ0}) and (\ref{definicaodecalQ1}) we will frequently
use the following functions
\begin{eqnarray}
  \calQ_2 (t) & := & \calQ_0(t) \int_0^t \left(
    \calQ_0(\tau)-M\left(\calQ_0\right)\right) d\tau ,
\label{definicaodecalQ2}
\\
\calQ_3 (t) & := & \calQ_0(t) \int_0^t \left(
  \calQ_1(\tau)-M\left(\calQ_1\right)\right) d\tau .
\label{definicaodecalQ3}
\end{eqnarray}
Note that, by their definitions, the functions $\calQ_i$, $i = 0, \,
\ldots, 3$, are quasi-periodic if $f$ is quasi-periodic.
We will have more to say about their properties below.


\section{Elimination of Secular Terms. The Main Results}
\label{sec:elimination} 
\zerarcounters

We recall in this section some of the methods and techniques
developed in the previous works~\cite{qp} and~\cite{pp}. A key
result for our method is the theorem below, proven in~\cite{qp},
which presents the solution of the Schr\"odinger equation
(\ref{equacaodeSchroedingerparaPhi}) in terms of particular
solutions of a generalised Riccati equation.

\begin{theorem} \label{teo:riccati}
Let $f: \R \to \R$, $f \in C^1 (\R)$ and $\eps \in \R$ and let $
g: \R \to \C$, $g \in C^1 (\R) $, be a particular solution of the
generalised Riccati equation
\begin{equation}
  \dot{G} -i G^2 - 2 i f G + i \eps^2  \; = \;  0 .
\label{eq:riccati}
\end{equation}
Then, the function $\Phi: \R \to \C^2$ given by
$
\Phi(t) = 
  \left(
       {\phi_+ (t) \atop \phi_- (t)}
  \right) =  U(t)\Phi(0) = U(t, \, 0)\Phi(0) ,
$
where
\begin{equation}
 U(t)  \; := \;
 \left(
  \begin{array}{cc}
     R(t)\left( 1 + i g(0) S(t)\right) & -i \eps R(t)S(t) \\
     &  \\
      -i \eps\overline{R(t)}\; \overline{S(t)} &
                        \overline{R(t)}\left(
                         1 - i\;\overline{g(0)}\; \overline{S(t)}\right)
    \end{array}
   \right) ,
\label{definicaodeU}
\end{equation}
with
\begin{equation*}
    R(t) \; := \; \exp\left( -i \int_0^t (f(\tau) + g(\tau))\,d\tau\right)
\quad \mbox{ and } \quad
    S(t) \; := \; \int_0^t R(\tau )^{-2} \; d\tau ,
\end{equation*}
is a solution of (\ref{equacaodeSchroedingerparaPhi}) with initial
value $ \Phi(0)  =
  \left(
      { \phi_+ (0) \atop \phi_- (0) }
  \right)  \in \C^2
$. $\EndofStatement\!\!$
\label{proposicaosobreaformadasolucaoemtermosdeg}
\end{theorem}

Let us briefly describe some of the ideas leading to Theorem
\ref{proposicaosobreaformadasolucaoemtermosdeg} and to other
results of \cite{qp}. As we saw in \cite{qp}, the solutions of the
Schr\"odinger equation (\ref{equacaodeSchroedingerparaPhi}) can be
studied in terms of the solutions of a particular complex version
of Hill's equation:
\begin{equation}
        \ddot{\phi}(t)  +  \left(  i \dot{f}(t) + \eps^2 +
         f(t)^2 \right)\phi (t)  \; = \;  0 .
\label{Hillprimeira}
\end{equation}
In fact, a simple computation shows that the components $ \Phi (t)
$ satisfy $ \ddot{\phi}_\pm + ( \pm i \dot{f} + \eps^2 + f^2
)\phi_+ = 0 $. If  we attempt to solve (\ref{Hillprimeira})
using the Ansatz
\begin{equation}
\phi (t) = \exp \left(-i\int_0^t (f(\tau)+g(\tau))
  d\tau\right) ,
\label{AnsatzparasolucaodeHill}
\end{equation}
it follows that $g$ has to satisfy the generalised Riccati
equation (\ref{eq:riccati}). We then try to find solutions for $g$
in terms of a power expansion in $\eps$ (vanishing for $\eps=0$) like
\begin{equation}
    g(t) \; = \; q(t)\sum_{n=1}^{\infty}v_n(t) \, \eps^n \, ,
\label{eq:ansatz}
\end{equation}
where the function $q$ was defined in (\ref{definicaodecalQ0}) and
is of central importance in this work.

The heuristic idea behind the Ans\"atze
(\ref{AnsatzparasolucaodeHill}) and (\ref{eq:ansatz}) is the
following. For $\eps \equiv 0$ a solution for (\ref{Hillprimeira}) is
given by $ \exp \left(-i\int_0^t f(\tau) d\tau\right)$. Thus, in
(\ref{AnsatzparasolucaodeHill}) and (\ref{eq:ansatz}) we are searching
for solutions in terms of an ``effective external field'' of the form
$ f+g$, with $ g$ given in terms of a convergent power series
expansion in $\eps$, vanishing for $\eps=0$.  A solution of
the form (\ref{AnsatzparasolucaodeHill}) leads to one of the two
independent solutions of (\ref{Hillprimeira}). The full solution of
(\ref{equacaodeSchroedingerparaPhi}) in terms of solutions of the
generalised Riccati equation (\ref{eq:riccati}) is that described in
Theorem \ref{proposicaosobreaformadasolucaoemtermosdeg} (see the
discussion of \cite{qp}).

We then proceed inserting (\ref{eq:ansatz}) into
(\ref{eq:riccati}). The result is a set of recursive first order
linear differential equations for the functions $v_n$ which can be
easily integrated. The solutions of these equations are
\begin{eqnarray}
\label{eq:v1} v_1(t) & = & \kappa_1 q(t) \, ,
\\
\label{eq:v2} v_2(t) & = & q(t) \left[ i \int_{0}^{t} (\kappa_1^2
\calQ_0(\tau) - \calQ_0(\tau)^{-1}) \, d\tau + \kappa_2 \right]  \, ,
\\
\label{eq:vn} v_n(t) & = & q(t) \left[ i \left(\int_{0}^{t}
\sum_{p=1}^{n-1} v_p(\tau) v_{n-p}(\tau) \, d\tau\right) +
\kappa_n \right]  \, , \;\;\;\; \text{for $n \geq 3$} \, ,
\end{eqnarray}
where the $\kappa_n$'s above, $n = 1, 2, \ldots$, are \underline{arbitrary}
integration constants. 
Defining 
$$
\calI_2(t) \; := \; \kappa_1^2 \calQ_0(t) - \calQ_0(t)^{-1}, 
\qquad 
\calI_n(t) \; := \sum_{p=1}^{n-1}v_p(t)v_{n-p}(t), 
\quad n \geq 3.
$$
we can write (\ref{eq:v1})-(\ref{eq:vn}) as
$$
v_1(t) \; = \; \kappa_1 q(t) , \qquad
v_n(t) \; = \; iq(t)\int_0^t\calI_{n}(\tau)d\tau + \kappa_n q(t), \quad n \geq 2.
$$
Observe that, in particular, we could just set all the $\kappa_n$'s
equal to zero. However, this is not a clever choice, since it would
result in polynomial terms on $t$ (the so-called {\it Secular Terms})
for the series expansion (\ref{eq:ansatz}) of $g$. This, of
course, would restrict the convergence of the series just for small
values of time.  As noticed in \cite{qp}, there is a choice of the
constants $\kappa_n$ for which one can eliminate completely all the
polynomial terms on $t$ that would eventually appear in $g$.  This
procedure, which we call the {\it Elimination of Secular Terms}, will
be briefly described now.

First of all, assuming that the function $f:\R \to \R$ is
quasi-periodic, it was proven in Appendix~B of~\cite{qp} that $q$,
defined in (\ref{eq:ansatz}), is also quasi-periodic.  Hence, $v_1$
in (\ref{eq:v1}) is quasi-periodic. The same is true for the integrand
$\calI_2$ which appears in $v_2$, equation (\ref{eq:v2}).
Recalling that $\calI_2$ depends on the free integration
constant $\kappa_1$, the key idea is to fix $\kappa_1$ in such a way
that the mean value of $\calI_2$ is equal to zero, that is
$
M(\calI_2)   =   M(\kappa_1^2 \calQ_0 - \calQ_0^{-1})   =   0 
$.
Since $\calQ_0$ is a quasi-periodic function, it readily
follows from this that
\begin{equation} \label{eq:kappa1fixado1}
\kappa_1^2  \; = \;  \frac{\overline{M(\calQ_0)}}{M(\calQ_0)} \, .
\end{equation}
With this choice of $\kappa_1$ one guarantees the absence of a
constant term in the Fourier expansion of $\calI_2$. Since
$\calI_2$ is being integrated in time, this would imply the
absence of a linear term on $t$ in the final expression for
$v_1$. An important remark is that (\ref{eq:kappa1fixado1})
will only make sense if we assume $M(\calQ_0) \neq 0$.

Under this assumption we can now proceed and fix recursively all
integration constants $\kappa_m$'s by \underline{imposing} a zero mean
value for the integrands $\calI_n$'s, $n = 3,\; 4,\; \ldots$, which
appear in (\ref{eq:vn}). This procedure removes, order by order in
$\eps$, the presence of the secular terms in the series expansion
(\ref{eq:ansatz}) for $g$ and recursively implies that all
functions $v_n$ are quasi-periodic.  Once all secular terms have been
removed, one can write the Fourier expansion for the functions
$v_n$ as
\begin{equation} \label{eq:fourierv}
v_n(t)  \; = \;  \sum_{\undm \in \Z^A} V^{(n)}_{\undm}  e^{i \undm
\cdot \undomega t} \, ,
\end{equation}
provided the sum converges absolutely. It was shown in~\cite{qp} that
this is indeed true. The proof of this fact was performed in the
following way: first it was shown that the Fourier coefficients
$Q_{\undm}$ of the function $q$ satisfies the bound $|Q_{\undm}|
\leq \calQ e^{-\chi |\undm|}$, for some $ \chi > 0$. Then, the method
of elimination of secular terms described above was applied to fix the
integration constants $\kappa_n$ leading to inductive bounds of the
form $|V^{(n)}_{\undm}| \leq \calK_n e^{-(\chi - \delta_n)|\undm|}$
for the Fourier coefficients of the function $v_n$, where $0 <
\delta_n < \chi$, for all $n = 1, 2, \ldots$. This exponential decay
is enough to proof the convergence of the sum in (\ref{eq:fourierv})
and to establish by induction the quasi-periodicity of all the
functions $v_n$.  Unfortunately, due to the bad behaviour in $n$ of
the constants $\calK_n$, it was not possible prove the convergence of
the $\eps$-expansion (\ref{eq:ansatz}), hence (\ref{eq:ansatz}) has to
be seen as a formal quasi-periodic power series solution of the generalised
Riccati equation (\ref{eq:riccati}).

The reason for the bad behaviour of $\calK_n$ is related to the
presence of convolutions and to the {\it small denominators} appearing
in the recursive relations for the coefficients $V_{\undm}^{(n)}$. A
general discussion of these problems is found in \cite{qp}.
However, in the situation where $f$ is a {\it periodic} function,
stronger results are possible.  In~\cite{pp}, where this situation was
studied, it was possible to prove the convergence of the power series
(\ref{eq:ansatz}) and uniform convergence of the Fourier series
involved in the computation of the wave functions. Moreover, absolute
convergence of the $\eps$-expansions leading to the {\it secular
  frequency} and to the coefficients of the Fourier expansion of the
wave functions was also proven.

All the work done in~\cite{qp} and~\cite{pp} was restricted to one of
the mutually exclusive conditions (I) and (II) of
Theorem~\ref{Teoremaprincipalsobreopropagador}. These conditions are
consequences of the method of elimination of secular terms. Clearly,
(I) is vital for (\ref{eq:kappa1fixado1}). When (I) is not satisfied
we have to apply condition (II). Both cases were studied in~\cite{qp}
and~\cite{pp}, where the method of elimination of secular terms has
been applied and equivalent results concerning the solution $g$ were
obtained. In the present work, we apply the method of elimination of
secular terms to study a more restrictive condition than those
represented by (I) and (II). Namely, we are concerned here with the
situation where $M(\calQ_0) = 0$ and also $M(\calQ_1) = 0$.

In this case, the complexity of the calculations involved to find the
right choice of constants $\kappa_n$'s grows enormously, in contrast
with those needed in the cases (I) and (II), already studied.  The
reason for that is quite simple: due to the hypothesis $M(\calQ_0) =
M(\calQ_1) = 0$ one needs to work explicitly with higher order terms
involved in the expansion (\ref{eq:ansatz}). This will become
more clear in Section~\ref{sec:case}.

In Section~\ref{sec:example} we will discuss an important example
where conditions (I) and (II) are not satisfied. To solve it, we have
to apply the solution (free of secular terms) obtained here. We present
in Section~\ref{sec:numerical} numerical calculations on this
particular example and obtained some interesting results.

We are ready now to state two main theorems of this work.

\begin{theorem} \label{teo:condIII}
Let $f:\R \to \R$ be a real quasi-periodic function satisfying 
\begin{center}
{\rm (III)} \, $M(\calQ_0) = M(\calQ_1) = 0$, 
but $M(\calQ_3) \neq 0$.
\end{center}
Then, there are constants $\kappa_n$, $n\geq 1$, such that all
functions $v_n$ given in (\ref{eq:v1})-(\ref{eq:vn}) are
\underline{quasi-periodic}. The explicit recursive expressions for the
constants $\kappa_n$ are found in (\ref{eq:capa123})-(\ref{eq:capan}).
$\EndofStatement$
\end{theorem}

The proof of this theorem is the main content of
Section~\ref{sec:case}. It states that the procedure of elimination of
secular terms outlined above also works under condition (III). When $f$ is
quasi-periodic this does not imply, however, that the formal solution
(\ref{eq:ansatz}) of the generalised Riccati equation
(\ref{eq:riccati}) converges, since we have the same difficulties
discussed in detail in~\cite{qp}.

For \underline{periodic} $f$, the situation is different and stronger
results can be proven. Let $f(t)=\sum_{m\in\Z}F_m
e^{i m \omega t}$ be a real periodic function with frequency $\omega$. If
$F_0=M(f)=0$, $q$ and $\calQ_0 = q^2$ are also periodic and their spectra of
frequencies are subsets of $\{n\omega , \; n\in \Z\}$. Following the
notation employed in \cite{pp}, we write the Fourier expansions of $q$
and $\calQ_0$ as
\begin{equation} \label{eq:notQ}
 q(t) \; = \; \sum_{m \in \Z} Q_{m} e^{i m \omega t},
\qquad
 \calQ_0(t) \; = \; q(t)^2 \; = \; \sum_{m\in \Z} Q_{m}^{(2)} e^{i m \omega t} .
\end{equation}
By relations (\ref{eq:v1})-(\ref{eq:vn}) and with the choice of
constants $\kappa_n$ mentioned in Theorem \ref{teo:condIII} (see
(\ref{eq:capa123})-(\ref{eq:capan})), the functions $v_n$ are also
periodic and their spectra of frequencies are also subsets of
$\{n\omega , \; n\in \Z\}$.  We write their Fourier expansions as
\begin{equation} 
v_n(t)  \; = \;  \sum_{m \in \Z} V^{(n)}_{m}  e^{i m \omega t}.
\label{eq:Vnm}
\end{equation}

In Appendix \ref{sec:convergence} we prove the theorem below, which
justifies our whole procedure for the case of \underline{periodic}
interactions and establishes convergence of (\ref{eq:ansatz}). 

\begin{theorem} \label{teo:convergence}
  Let $f(t)$ be as above with $F_0=M(f)=0$ and such that condition
  (III) of Theorem \ref{teo:condIII} is satisfied.  Moreover, assume
  that the coefficients $Q_{m}$ and $Q_{m}^{(2)}$ above satisfy the
  following: for any $ \chi >0$ there is a positive constant $ \calQ
  \equiv \calQ(\chi)$ such that
\begin{equation} %
\left|Q_m\right| \; \leq \; \calQ \frac{e^{-\chi |m|}}{\novomod{m}^2}
\quad
\mbox{ and }
\quad
      \left|Q_m^{(2)}\right| \; \leq \; \calQ \frac{e^{-\chi |m|}}{\novomod{m}^2} ,
\label{RR1}
\end{equation} %
for all $ m\in \Z$, where the symbol $\novomod{m} $ was defined in
(\ref{definicaodenovomod}).  Then, with the constants $\kappa_n$ fixed
as in Theorem \ref{teo:condIII}, the Fourier coefficients of the
functions $v_n$ given in (\ref{eq:v1})-(\ref{eq:vn}) satisfy
$$
\left| V^{(n)}_{m}\right| \; \leq \; 
M_0\, (M_1)^n \;\frac{e^{-\chi |m|}}{\novomod{m}^2}
$$
for all $m\in \Z$ and all $n\geq 1$, for some positive constants
$M_0$ and $M_1$. As a consequence, the power series expansion
(\ref{eq:ansatz}), representing a solution of the generalised Riccati
equation (\ref{eq:riccati}), converges uniformly for $t\in \R$,
provided $|\eps| < 1/M_1$. $\EndofStatement$
\end{theorem}

\begin{remark} 
  In~\cite{qp}, (\ref{RR1}) was established for $f$ is periodic and
  represented by a \underline{finite} Fourier series.  The condition
  $F_0=M(f)=0$ above is not crucial and can be eliminated following
  the procedure described in~\cite{pp}.
\end{remark}

It follows from this theorem that the main consequences of
Theorem~\ref{teo:floquet} are valid under condition (III) as well. In
particular, the Floquet form (\ref{U11U12}) holds and the secular
frequency $\Omega$ and the Fourier coefficients of (\ref{u11u12}) are
analytic functions of $\eps$ for $|\eps|$ small enough.
See Appendix \ref{app:novo} for a comment on this.

\subsection{The Secular Frequency}
\label{sec:secfreq} 

A feature of our method is that it allows to present the complete
$\epsilon$-expansion for the secular frequency $\Omega$ (also known as
Rabi frequency) associated to the solutions of
(\ref{equacaodeSchroedingerparaPsi})-(\ref{equacaodeSchroedingerparaPhi})
(see (\ref{U11U12})). One has (see \cite{qp,pp,BarataWreszinski2})
$$
\Omega \; = \; M(f)+M(g)= F_0 + \sum_{n=1}^{\infty} \epsilon^n
M\left(qv_n\right)\,.
$$
By Theorem~\ref{teo:convergence} above, this expansion is
convergent for $|\eps|$ small enough.
The knowledge of the complete expansion is particularly important for
the qualitative investigation of the large-time behaviour of that
solutions. After some simple calculations using
(\ref{eq:v1})-(\ref{eq:vn}) one gets
\begin{eqnarray}\label{eq:Omegacasogeral}
\Omega & = & 
F_0 + \epsilon\kappa_1M(\calQ_0) 
+\epsilon^2\left[i\kappa_1^2M(\calQ_2) -iM(\calQ_1) + 
\kappa_2M(\calQ_0)\right] \nonumber
\\ & &
+ \, \epsilon^3\left[
2\kappa_1M(\calQ_3) + \kappa_3 M(\calQ_0)
\right]
+O(\epsilon^4)\, . 
\end{eqnarray}
As we will show in Corollary \ref{corol:MQ2equal0} below,
$M(\calQ_0)=0$ implies $M(\calQ_2)=0$. Hence, for case (II),
\begin{equation}\label{eq:OmegacasoII}
\Omega \; = \; F_0 -i\epsilon^2M(\calQ_1) + 2\kappa_1 \epsilon^3
M(\calQ_3) +O(\epsilon^4)\, ,
\end{equation}
and for case (III),
\begin{equation}\label{eq:OmegacasoIII}
\Omega \; = \; F_0 + 2\kappa_1 \epsilon^3 M(\calQ_3) +O(\epsilon^4)\, .
\end{equation}
Actually, after fixing $\kappa_1$ in Section
\ref{sec:fixingkappa1}, we will see that $\Omega = F_0 + 2\epsilon^3
\left|M(\calQ_3)\right| +O(\epsilon^4)$.

In this case (III), if one additionally has $F_0 =0$, then $\Omega =
O(\eps^3)$, a fact first pointed in \cite{BarataWreszinski2}.  This
implies long transition times for certain probability amplitudes, a
phenomenon known as (approximate) dynamical localisation (see
\cite{Sacchetti} and other references therein). In Sections
\ref{sec:example} and \ref{sec:numerical} we discuss this situation
for $f$ describing a monochromatic interaction.


\section{Properties of the Mean Value and the Renormalisation Operator}
\label{sec:PMVRP} 
\zerarcounters

Let us introduce some notations that will be very useful. 
Since expressions like $f(\xi) \int_0^\xi d\xi' \, g(\xi')$ will
often appear throughout our calculations, we define a shorthand
notation
\begin{equation} \label{eq:short}
\q{f}{g}_\xi \defi f(\xi) \int_0^\xi d\xi' \, g(\xi') \, .
\end{equation}
Moreover, if $\q{f}{g}_t$ is quasi-periodic function on $t$, then
$M\q{f}{g}$ will denote its mean value. We also define
\begin{equation}
\qq{f}{g}{h}_\xi \defi \q{f}{\q{g}{h}_{\xi'}}_\xi \; = \; f(\xi)
\int_0^\xi d\xi' \, g(\xi') \int_0^{\xi'} d\xi'' \, h(\xi'') \, .
\end{equation}
Further compositions like $( f_1\,|f_2\,|\,\cdots\,|\,f_n )_\xi$
are defined in analogous way, so that, for $n>2$,  
$$
( f_1\,|f_2\,|\,\cdots\,|\,f_n )_\xi \; := \; 
\q{f_1}{\left( f_2\,|\,\cdots\,|\,f_n \right)_{\xi'}}_\xi.
$$ 
Note, finally, the trivial fact that 
\begin{equation}
f_0(\xi)( f_1\,|f_2\,|\,\cdots\,|\,f_n )_\xi \; = \; 
( f_0f_1\,|f_2\,|\,\cdots\,|\,f_n )_\xi  \, .
\label{eq:oiubnpsS}
\end{equation}

\subsection{Properties of the Mean Value} 
\label{sec:mean}

The following general results on the mean value of some quasi-periodic
functions (see definition (\ref{eq:media})) will be used for many
purposes in the present work.

\begin{proposition} \label{prop:1}
Let $a: \R \to \C$, $b: \R \to \C$ and $c: \R \to \C$ be
quasi-periodic functions with Fourier components denoted by 
$A_{\undm}$, $B_{\undm}$ and $C_{\undm}$, $\undm \in \Z^A$,
respectively. We have the following statements:
\begin{enumerate}
\item[1.] If $M(c) = 0$, then the Fourier components of the function
$h(t):=\q{b}{c}_t$ are given by
\begin{equation} \label{eq:decomp}
H_{\undn}  \; = \;  \sum_{\undm \in \Z^A_\ast} \frac{i \,
C_{\undm} \left( B_{\undn} - B_{\undn - \undm} \right)}{\undm
\cdot \undomega} \, ,
\end{equation}
for all $\undn \in \Z^A$. Moreover, if $M(b) = 0$ and
$C_{\undm}B_{-\undm} = C_{-\undm} B_{\undm}$ for all $\undm \in
\Z^A_{\ast}$, then $M(h) = 0$.
\item[2.] If $M(b) = 0$ and $M(c) = 0$, then $M\q{b}{c} =
-M\q{c}{b}$.
\item[3.] If $M(c) = 0$, then $M\q{c}{c} = 0$.
\item[4.] If $M(a) = M(c) = M\q{b}{c} = 0$, then
\begin{equation}
M\qq{a}{b}{c} \; = \; - M\left[b(t) \left(\int_0^t a(\tau) \,
d\tau \right)\left(\int_0^t c(\tau) \, d\tau \right) \right] \, . 
\label{eq:ineirbv}
\end{equation} 
Moreover, if also $M\q{b}{a} = 0$, then $M\qq{a}{b}{c} =
M\qq{c}{b}{a}$.
\item[5.] If $M(a) = M(b) = M(c) = 0$ and $M\q{b}{c} = M\q{c}{a} =
M\q{a}{b} = 0$, then
\begin{equation}
M\qq{a}{b}{c} + M\qq{b}{c}{a} + M\qq{c}{a}{b} \; = \; 0 \, .
\label{eq:remarkable}
\end{equation}
\item[6.] If $M(a)=0$, it follows from (3) and (5) that $M\qq{a}{a}{a} = 0$.
$\EndofStatement$
\end{enumerate}
\end{proposition}

The identity (\ref{eq:remarkable}) is very remarkable. We have noticed
that some non-trivial relations follow from it, specially if one write
(\ref{eq:remarkable}) in terms of the Fourier coefficients of $a$, $b$
and $c$.

\vspace{0.2cm}
\Proof 
We can demonstrate (1) by explicitly computing the Fourier
decomposition of $h(t)$. Thus,
\begin{eqnarray*}
h(t)  &=&  \sum_{\undn \in \Z^A} B_{\undn} e^{i \undn \cdot
\undomega t} \left[ \sum_{\undm \in \Z^A_\ast} \frac{C_{\undm}}{i
\undm \cdot \undomega} \left( e^{i \undm \cdot \undomega t} - 1
\right) \right] \\ &=& \sum_{\undn \in \Z^A} \left[ \sum_{\undm
\in \Z^A_\ast} \frac{i \, C_{\undm} \left( B_{\undn} - B_{\undn -
\undm} \right) }{\undm \cdot \undomega} \right] e^{i \undn \cdot
\undomega t} \, ,
\end{eqnarray*}
proving the first statement of (1). Now, if $M(b) = B_0 = 0$ and
$C_{\undm}B_{-\undm} = C_{-\undm} B_{\undm}$ for all $\undm \in
\Z^A_{\ast}$, then
\begin{equation} \label{eq:prova2}
H_0  \; = \;  \sum_{\undm \in \Z^A_\ast} \frac{- i \, C_{\undm}
B_{-\undm}}{\undm \cdot \undomega}  \; = \;  
\frac{-i}{2}\sum_{\undm \in
\Z^A_{\ast }} \frac{ (C_{\undm}B_{-\undm} -
C_{-\undm}B_{\undm})}{\undm \cdot \undomega}  \; = \;  0 \, .
\end{equation}
Since $M(h) = H_0$, we completed the proof of (1).
To demonstrate (2) we simply use the first equality of
(\ref{eq:prova2}) to write
\begin{equation*}
M\q{b}{c}  \; = \;  \sum_{\undm \in \Z^A_\ast} \frac{- i \, C_{\undm}
B_{-\undm}}{\undm \cdot \undomega} \hspace{0.5cm} \text{ and }
\hspace{0.5cm} M\q{c}{b}  \; = \;  \sum_{\undm \in \Z^A_\ast} \frac{-
i \, B_{\undm} C_{-\undm}}{\undm \cdot \undomega} \, .
\end{equation*}
Changing $\und{m} \to -\und{m}$ in the second equation above, we get
the desired claim.

Statement (3) is a mere consequence of (2) when we take $b = c$.
Statement (4) can be proven using (2). Indeed, let $h(t) :=
\q{b}{c}_t$. Since $M(h) = M\q{b}{c} = 0$ and $M(a) = 0$, by (2), 
we can write
\begin{eqnarray*}
M\q{a}{h} \;\;=\;\; - M\q{h}{a} &=& -M\left(h(t) \int_0^t 
a(\tau) \, d\tau \right) \\ &=& - M \left[b(t)\int_0^t c(\tau) 
\, d\tau \left(\int_0^t a(\tau) \, d\tau \right) \right] \, ,
\end{eqnarray*}
which proves the first claim. To prove the second one, all we need to
do is to interchange the the roles of $c$ and $a$ in the last equality
(note that since $M\q{b}{a} = 0$, the mean value of $\qq{c}{b}{a}_t$
is well defined).  Finally, statement (5) can be easily obtained as
follows.  Using (\ref{eq:ineirbv}), writing
$b(t)=\frac{d}{dt}\int_0^tb(\tau)d\tau$, using definition
(\ref{eq:media}) and integration by parts, one gets
\begin{eqnarray*} \label{eq:integralabc}
M\qq{a}{b}{c} &=& - \lim_{T\to\infinito} \frac{1}{2T} 
\left. \left(\int_0^t b(\tau) \, d\tau\right) 
\left(\int_0^t a(\tau) \, d\tau \right) \left(\int_0^t c(\tau) \,
d\tau \right)\right|_{-T}^{T} \\ & & + M\left(\left(\int_0^t b(\tau) \, d\tau\right) 
a(t) \, \left(\int_0^t c(\tau) \, d\tau
\right) \right) \\ & & + M\left(\left(\int_0^t b(\tau) \,
d\tau \right) \left(\int_0^t a(\tau) \, d\tau \right) 
\, c(t)  \right) \, .
\end{eqnarray*}
Now, since $M(a) = M(b) = M(c) = 0$, the integrals in the first line
of (\ref{eq:integralabc}) are all bounded. Thus, the limit $T \to
\infinito$ is zero because of the division by $2T$. Applying
(\ref{eq:ineirbv}) to the remaining terms we obtain (5). \QED

The following trivial corollary is of crucial importance for some of our
calculations:
\begin{corollary}
For $M(\calQ_0)=0$ one always has $M(\calQ_2)=0$. $\EndofStatement\!\!\!$
\label{corol:MQ2equal0}
\end{corollary}

\vspace{0.2cm}
\Proof If $M(\calQ_0)=0$ then, by (\ref{definicaodecalQ2}), $\calQ_2
(t) = \q{\calQ_0}{\calQ_0}_t$. Hence, from statement (3) of
Proposition~\ref{prop:1}, it follows that $M(\calQ_2)=0$. \QED

\subsection{The Renormalisation Operator}
\label{sec:RenOp}

For general quasi-periodic functions $a_1, \; \ldots, \; a_n$, the
function $(a_1 \,|\, \cdots \,|\, a_n)_t$, defined above, is not
generally quasi-periodic, since an integration performed on a
quasi-periodic function with a non-zero mean value would produce a
(linear in $t$) secular term, which would eventually become a higher
degree polynomial after further integrations. We will here describe an
operation designed to produce a quasi-periodic function out of $(a_1
\,|\, \cdots \,|\, a_n)_t$ through interactive subtractions of the
mean value of the functions being integrated, a procedure we call
``renormalisation'' due to the analogy to the procedure of
perturbative renormalization in quantum field theory.  We will use
this procedure of renormalisation in the following sections and here
we present its definition and basic properties.

Let $a_1, \; \ldots, \; a_n$ be quasi-periodic functions.  We define
inductively the {\it renormalisation operator} $\ren{n}$ acting on
$(a_1 \,|\, \cdots \,|\, a_n)$ by
\begin{eqnarray*}
\ren{1} a_1 (t) &:= & a_1(t), \\
\ren{2}\q{a_1}{a_2}_t & := & \q{a_1}{\ren{1}(a_2)-M(\ren{1}(a_2))}_t 
                     \; = \; \q{a_1}{a_2-M(a_2) }_t , \\
\ren{n}(a_1 \,|\, \cdots \,|\, a_n)_t
 &:= & \q{a_1}{\, \ren{n-1}(a_2 \,|\, \cdots \,|\, a_{n}) 
                 - M(\ren{n-1}(a_2 \,|\, \cdots \,|\, a_{n}))\, }_t,
\end{eqnarray*}
for $n > 2$. We will now prove some elementary facts on $\ren{n}$ which will be
used below. The first important observation is that if $a_1, \;
\ldots, \; a_n$ are quasi-periodic functions, then $\ren{n}(a_1 \,|\,
\cdots \,|\, a_n)$ is also quasi-periodic. This can be easily seen by
induction, through the obvious remark that the mean value of
$\ren{n-1}(a_2 \,|\, \cdots \,|\, a_{n}) - M(\ren{n-1}(a_2 \,|\,
\cdots \,|\, a_{n}))$ is zero.
Note also that, trivially
\begin{equation}
a_0\ren{n}(a_1 \,|\, \cdots \,|\, a_{n})
\; = \; 
\ren{n}(a_0a_1 \,|\, \cdots \,|\, a_{n}) \, .
\label{eq:bvsiupi}
\end{equation}

The following proposition is a trivial but useful restatement of the
definition of the $\ren{n}$'s:

\begin{proposition}
  For all $n\geq 2$ the following statement holds: if $a_1, \; \ldots,
  \; a_n$ are quasi-periodic functions, then
\begin{equation*}
\ren{n} (a_1 \,|\, \cdots \,|\, a_n) \; = \;
\ren{2} \q{a_1}{\ren{n-1} (a_2 \,|\, \cdots \,|\, a_n)} .
\end{equation*}
Consequently, for $n>2$,
\begin{equation}
\ren{n} (a_1 \,|\, \cdots \,|\, a_n) \; = \;
\ren{2} \q{a_1}{\, \ren{2}\q{a_2}{
                           \ren{2}(\, \cdots \, |\ren{2}\q{a_{n-1}}{a_n}}\cdots)\, 
                               } .
\label{eq:ionviPOI}
\end{equation}
$\EndofStatement\!\!$
\label{prop:SSH1}
\end{proposition}

\vspace{-0.3cm}

\Proof By the definition of $\ren{2}$,
\begin{eqnarray*}
\hspace{0.5cm}
\ren{2} ( a_1 \,| \ren{n-1} (a_2 \,| \!\!\!\!&\cdots&\!\!\!\! |\, a_n)) =   
\\ &=& \q{a_1}{\;\ren{n-1} (a_2 \,|\, \cdots \,|\, a_n) - M(\ren{n-1} (a_2 \,|\,
\cdots \,|\, a_n))\;} \\ &=& 
\ren{n} (a_1 \,|\, \cdots \,|\, a_n) . \hspace{5.6cm} \rule{1.8mm}{1.8mm}
\end{eqnarray*}

Relation (\ref{eq:ionviPOI}) shows that the operation $\ren{n}$ can be
obtained by iteration of the operation $\ren{2}$. One also has the
following useful

\begin{proposition}
  For all $n\geq 1$ the following statement holds: if $a_1, \; \ldots,
  \; a_n$ are quasi-periodic functions and $a_n = \ren{2}\q{b}{c}$ for
  quasi-periodic functions $b$, $c$, then
\begin{equation*}
  \hspace{2.2cm}
  \ren{n}(a_1 \,|\, \cdots \,|\, a_{n-1}\,|\, a_n) \; = \;
  \ren{n+1}(a_1 \,|\, \cdots \,|\, a_{n-1}\,|\, b\,|\, c).
  \hspace{1.9cm} \Box
\end{equation*}
\label{prop:SSH2}
\end{proposition}

\vspace{-0.5cm}

\Proof For $n=1$, let $a_1=\ren{2}\q{b}{c}$. Then $\ren{1} a_1 =a_1 =
\ren{2}\q{b}{c}$, trivially.  For $n=2$, let $a_2=\ren{2}\q{b}{c}$.
Then, $ \ren{2}\q{a_1}{a_2} = \q{a_1}{a_2-M(a_2)} = \q{a_1}{ \ren{2}
  \q{b}{c} - M(\ren{2}\q{b}{c}) } , $ but, by definition,
$$
\ren{3}\qq{a_1}{b}{c} = \q{a_1}{ \ren{2}\q{b}{c} - M(\ren{2}\q{b}{c})}
$$ 
and the statement holds again. For $n>2$, let
$a_n=\ren{2}\q{b}{c}$. Then, by induction,
\begin{eqnarray*}
\ren{n+1}(a_1 \,| \!\!\!\!&\cdots&\!\!\!\! |\, a_{n-1}\,|\, b\,|\, c) 
 = \\
&=& \q{a_1}{\;\ren{n}(a_2 \,|\, \cdots \,|\, a_{n-1}\,|\, b\,|\, c) 
-
M(\ren{n}(a_2 \,|\, \cdots \,|\, a_{n-1}\,|\, b\,|\, c) )\;} 
\\ 
&=& \q{a_1}{\;\ren{n-1}(a_2 \,|\, \cdots \,|\, a_{n-1}\,|\, a_n) 
-
M(\ren{n-1}(a_2 \,|\, \cdots \,|\, a_{n-1}\,|\, a_n )\;}
\\ 
&=& \ren{n}(a_1 \,|\, \cdots \,|\, a_{n-1}\,|\, a_n). 
\hspace{6.4cm} \rule{1.8mm}{1.8mm}
\end{eqnarray*}

If $a$ and $b_1, \; \ldots, \; b_m$ are quasi-periodic, a function
like $ \sum_{k=1}^m \q{a}{ b_k} $ may not be a sum of quasi-periodic
functions, even when $ \q{a}{ \sum_{k=1}^mb_k}= \sum_{k=1}^m\q{a}{b_k}
$ is quasi-periodic, since we are not assuming that $M(b_k)=0$ for
each individual $k$.  This fact notwithstanding, the following simple
statement holds and will be repeatedly used:

\begin{proposition}
  Let $a$ and $b_1, \; \ldots, \; b_m$  be
  quasi-periodic functions. Then
\begin{equation}
\ren{2}\q{a}{\sum_{k=1}^m b_k} 
\; = \; 
\sum_{k=1}^m\ren{2}\q{a}{b_k} .
\label{eq:IubuIssd2fg}
\end{equation}
Consequently, 
\begin{equation}
\ren{n}
\left(
a_1 \, \left| \, \cdots \,   \left| \, a_{n-1} \,   \left|\,
\sum_{k=1}^m b_k 
\right.\right.\right.
\right)
\; = \; 
\sum_{k=1}^m\ren{n} (a_1 \,|\, \cdots \,|\, a_{n-1}\,|\, b_k ) 
\label{eq:aincuib}
\end{equation}
for quasi-periodic functions $a_1, \; \ldots, \; a_{n-1}$ and $b_1, \; \ldots,
\; b_m$.  $\EndofStatement\!\!$
\label{prop:SSH3}
\end{proposition}

\vspace{0.2cm}
\Proof 
We have
\begin{eqnarray*}
\ren{2}\q{a}{\sum_{k=1}^m b_k} 
& = & 
\q{a}{\sum_{k=1}^m b_k  - 
           M\left(\sum_{k=1}^m b_k\right) } 
\; = \; 
\q{a}{\sum_{k=1}^m\left( b_k  - 
           M\left( b_k\right)\right) } 
\\
& = &
\sum_{k=1}^m \q{a}{ b_k  - M ( b_k ) } 
\; = \;  
\sum_{k=1}^m \ren{2} \q{a}{ b_k} .
\end{eqnarray*}
Note, in the third equality, that the mean value of $b_k - M ( b_k )$
is zero and, hence, $\q{a}{ b_k - M ( b_k ) }$ are quasi-periodic.
Relation (\ref{eq:aincuib}) follows from
(\ref{eq:ionviPOI})-(\ref{eq:IubuIssd2fg}). \QED

The following corollary follows from Propositions 
\ref{prop:SSH2} and \ref{prop:SSH3}.
\begin{corollary}
  Let  $a_1, \; \ldots, \;a_{n-1}$, $b_1, \; \ldots, \;b_m$ and $c_1, \;
  \ldots, \; c_m$ be
  quasi-periodic functions. Then,
\begin{equation*}
\hspace{0.2cm}
\ren{n}
\left(
a_1 \, \left| \, \cdots \,   \left| \, a_{n-1} \,   \left|\,
\sum_{k=1}^m \ren{2}\q{b_k}{c_k} 
\right.\right.\right.
\right)
\; = \; 
\sum_{k=1}^m\ren{n+1} (a_1 \,|\, \cdots \,|\, a_{n-1}\,|\, b_k\,|\, c_k ). 
\hspace{0.40cm}
\Box
\end{equation*}
\end{corollary}

\begin{remark}
  The reader should be warned of the following fact: If $a_1, \;
  \ldots, \; a_n$, $b_1, \; \ldots, \;b_m$ are quasi-periodic
  functions and $ (a_1 \,|\, \cdots \,|\, a_{n}) + (b_1 \,|\, \cdots
  \,|\, b_{m}) $ is also quasi-periodic, it is not always true that $
  (a_1 \,|\, \cdots \,|\, a_{n}) + (b_1 \,|\, \cdots \,|\, b_{m})$
  equals 
  $$ 
  \ren{n}(a_1 \,|\, \cdots \,|\, a_{n}) + \ren{m}(b_1 \,|\,
  \cdots \,|\, b_{m}). 
  $$  
  For a counter-example, take $a_1(t) = 1$,
  $a_2(t)= e^{it}$, $a_3(t) =1$ $b_1(t) = i e^{it}$ and $b_2(t)=1$.
  One has $\qq{a_1}{a_2}{a_3}+\q{b_1}{b_2}=e^{it}-1$, but
  $\ren{3}\qq{a_1}{a_2}{a_3}+\ren{2}\q{b_1}{b_2}=0$.  Hence, the
  renormalisation operations are not additive in this sense.
\end{remark}

\subsection{More on the Notation and Some Definitions}

With the shorthand notation introduced in (\ref{eq:short}) and the
definition of $\calQ_0$ in (\ref{definicaodecalQ0}), we see from
(\ref{definicaodecalQ1}), (\ref{definicaodecalQ2}) and
(\ref{definicaodecalQ3}) that
\begin{equation*} 
\calQ_1(t) \; = \; \ren{2}\q{\calQ_0}{\overline{\calQ_0}}_t,
\;\;\;
\calQ_2(t) \; = \; \ren{2}\q{\calQ_0}{\calQ_0}_t,
\;\;\;
\calQ_3(t) \; = \; \ren{2}\q{\calQ_0}{\calQ_1}_t \, . 
\end{equation*}
Below, we will often use the following compact notation
$$
\q{i}{j}_t \, :=\, \q{\calQ_i}{\calQ_j}_t , \quad
\q{i}{\overline{j}}_t \, :=\, \q{\calQ_i}{\overline{\calQ_j}}_t ,
\quad \qq{i}{j}{k}_t \, :=\, \qq{\calQ_i}{\calQ_j}{\calQ_k}_t,
$$
etc, for $i, \; j, \; k= 0,\, \ldots,\, 3$.  In other words, we
simply use the index $n$ of $\calQ_n$ to denote $\calQ_n$ itself.
Moreover, by $M\q{i}{j}$ we will denote the mean value of
$\q{i}{j}_t$, etc.

Note that for $M(\calQ_0)=0$ one has with this notation
$
\calQ_1(t)  =  \q{0}{\overline{0}}_t
$ and
$
\calQ_2(t)  =  \q{0}{0}_t 
$
and for $M(\calQ_1)=0$ one has 
$$
\calQ_3(t) \; =\;  \q{0}{1}_t .
$$
We will often write $\calQ_3$ this way.

For $M(\calQ_0)=0$, other identities will also be at hand. For instance, one has
\begin{equation}
\qq{0}{0}{0}_t \; = \; \q{0}{\q{0}{0}_{t'}}_t  \; = \; \q{0}{2}_t 
\quad
\mbox{ and }
\quad
\q{1}{0}_t \; = \; \q{2}{\overline{0}}_t \, , 
\label{eq:iduteis}
\end{equation}
since $M\q{0}{0}=0$, by item (3) of Proposition \ref{prop:1}.
Relations like these will be often employed.


\section{The Case $M(\calQ_0) = 0$ and $M(\calQ_1) = 0$}
\label{sec:case} 
\zerarcounters

In this Section we will prove the Theorem \ref{teo:condIII}.  Our
interest is to study the situation complementary to cases (I) and
(II), i.e., the situation where one has condition
\begin{center}
($\mbox{III}_0$) $\; M(\calQ_0) = 0$ and $M(\calQ_1) = 0$.
\end{center}
To remove the secular terms from $g$, applying the method
described in the previous section, we will be forced to add a further
restriction to ($\mbox{III}_0$), namely the condition $M\q{0}{1}\neq 0$. 

Recall that the functions $q$ and $\calQ_1$ depend primordially
on the interaction $f$ (see the definitions given in
(\ref{definicaodecalQ0}) and (\ref{definicaodecalQ1})), so conditions
(I), (II) or ($\mbox{III}_0$) apply upon the properties of $f$.  As we
already saw in Section~\ref{sec:intro}, the function $f(t) = \varphi_1
\cos(\omega t) + \varphi_2\sin(\omega t)$ only satisfies condition (I)
or ($\mbox{III}_0$), depending on the particular choice of the
parameters $\varphi_1, \varphi_2$. We will have more to say about this
example latter on in Section~\ref{sec:example}. Now, let us work with
the expansion for $g$ in order to remove all of its secular terms.

Again, our Ansatz to solve the generalised Riccati equation
(\ref{eq:riccati}) is (\ref{eq:ansatz}). The explicit solutions for
the coefficients $v_n$ are given in (\ref{eq:v1}), (\ref{eq:v2}) and
(\ref{eq:vn}). One sees immediately from condition ($\mbox{III}_0$)
that $v_1$ and $v_2$ do not suffer from secular terms. Indeed, $q$
in quasi-periodic and since $M(\calQ_0^{-1}) = \overline{M(\calQ_0)} =
0$, we conclude that the mean value of the integrand $\calI_2$
occurring in (\ref{eq:v2}) is zero. Therefore, the integration
occurring in the definition of $v_2$ in (\ref{eq:v2}) does not produce
a linear term in $t$. These fats imply that $v_1$ and $v_2$ are
quasi-periodic under ($\mbox{III}_0$).  From these considerations we
see that the condition $M(\calI_n)=0$, $n\geq 3$, becomes recursively
identical to $\sum_{p=1}^{n-1}M(v_pv_{n-p})$, $n\geq 3$, since the
$v_n$'s become successively quasi-periodic when the interactive
procedure is run.

If this is achieved, i.e., if we succeed in fixing $M(\calI_n)=0$ for
all $n$, we can rewrite (\ref{eq:v1})-(\ref{eq:vn}) in a
``renormalised'' form: 
\begin{eqnarray}
v_1(t) & = & \kappa_1 q(t) , \label{eq:v1ren}\\
v_2(t) & = & q(t)^{-1}\left(i\kappa_1^2\calQ_2(t) -i \calQ_1(t) +
  \kappa_2 \calQ_0(t) \right) ,
\label{eq:v2ren}\\
v_n(t) & = &
q(t)^{-1}
\left\{
i\sum_{p=1}^{n-1}\ren{2}\q{0}{v_pv_{n-p}}_t + \kappa_n \calQ_0(t) 
\right\}, \quad n \geq 3,
\label{eq:vnren}
\end{eqnarray}
where $\calQ_0$, $\calQ_1$ and $\calQ_2$ were defined in
(\ref{definicaodecalQ0}), (\ref{definicaodecalQ1}) and (\ref{definicaodecalQ2}),
respectively.

Let us move on and analyse the third order term. According to
(\ref{eq:vn}) the integrand $\calI_3$ which appears in the definition
of $v_3$ is given by $2v_1v_2$.  Using (\ref{eq:v1ren}) and
(\ref{eq:v2ren}) we have
\begin{eqnarray} 
\label{eq:2v1v2}
v_1 v_2 \; = \; i\kappa_1^3 \calQ_2 - i\kappa_1 \calQ_1 +
\kappa_1\kappa_2 \calQ_0 \, .
\end{eqnarray}
From Corollary \ref{corol:MQ2equal0}, we readily see that
$M(\calI_3)=2M(v_1v_2) = 0$.  This means that the integrand which
appears in $v_3$ does not have a constant term in its Fourier
expansion. Hence, $v_3$ is quasi-periodic.

Until now we have verified the absence of secular terms in the
series expansion of $g$ up to order three in $\eps$. As we shall see
next, for the same to be true up to order four, we have to
make a especial choice for the value of the constant $\kappa_1$.

\subsection{The Absence of Secular Terms in $v_4$. Fixing
$\kappa_1$}
\label{sec:fixingkappa1}

As one sees from (\ref{eq:vn}), the integrand in $v_4$ is $\calI_4
\defi 2v_1v_3 + v_2^2$. Since $v_1, v_2, v_3$ are quasi-periodic, the
mean value of $\calI_4$ is well defined.  Let us explicitly evaluate
$\calI_4$ using (\ref{eq:v1})--(\ref{eq:vn}):
\begin{eqnarray}
\calI_4 &=& 2 \kappa_1 \calQ_0(t) \left(i\int_0^t
2v_1(\tau)v_2(\tau) \, d\tau + \kappa_3 \right)  
\nonumber \\ & & + \; 
\calQ_0(t)\left(i\int_0^t\left(\kappa_1^2 \calQ_0(\tau) - \calQ_0(\tau)^{-1}\right) \,
d\tau + \kappa_2\right)^2
\nonumber \\
&=& 
-4\kappa_1^4 \q{0}{2} + 4\kappa_1^2 \q{0}{1} + 6i
\kappa_1^2\kappa_2 \calQ_2 + (2 \kappa_1\kappa_3 + \kappa_2^2)
\calQ_0
\nonumber \\
& & - 2 i \kappa_2 \calQ_1 - \kappa_1^4 \q{2}{0} +
2\kappa_1^2 \q{2}{\bar{0}} -\q{1}{\bar{0}} \, . \label{eq:desI4}
\end{eqnarray}

The functions appearing in the right-hand side of (\ref{eq:desI4}) are
all quasi-periodic, since $M(\calQ_0) = M(\calQ_1) = M(\calQ_2) = 0$.
Therefore, we are allowed to take the mean value of each individual
term above. The result is
\begin{equation*} %
M(\calI_4) \, = \, -4\kappa_1^4 M(\,0\,|\,2\,) + 4\kappa_1^2 M(\,0\,|\,1\,)  -
\kappa_1^4 M(\,2\,|\,0\,) + 2 \kappa_1^2 M(\,2|\,\bar{0}\,) - M(\,1\,|\,\bar{0}\,).
\end{equation*}
By statements (2) and (6) of Proposition~\ref{prop:1} and by
(\ref{eq:iduteis}) we have 
\begin{eqnarray} 
\label{eq:M1}
M\q{2}{0} & = & -M\q{0}{2} \; = \; -M\qq{0}{0}{0} = 0 \, ,
\\
\label{eq:M2}
M\q{2}{\bar{0}} 
& = & 
M\q{1}{0} 
\; = \; 
-M\q{0}{1} 
\, .
\end{eqnarray}
Hence,
\begin{equation} \label{eq:I4}
M(\calI_4) \; = \; 2\kappa_1^2 M\q{0}{1}  
- M\q{1}{\bar{0}} \, . 
\end{equation} 
We have the following
\begin{proposition} \label{prop:III}
Under $M(\calQ_0)=M(\calQ_1)=0$, one has 
$M(\,1\,|\,\bar{0}\,) = 2 \overline{M(\,0\,|\,1\,)}.$~$\EndofStatement$
\end{proposition}
\Proof By the definition of $\calQ_1$,
\begin{equation} \label{eq:10}
M\q{0}{1} \;=\; M\qq{0}{0}{\bar{0}} 
\end{equation}
and
\begin{equation} \label{eq:1b0}
M\q{1}{\bar{0}} = M\left(\calQ_0(t) \left(\int_0^t \calQ_0(\tau)^{-1} \, d\tau
\right) \left(\int_0^t \calQ_0(\tau)^{-1} \, d\tau\right)\right) =
-M\qq{\bar{0}}{0}{\bar{0}} \, ,
\end{equation}
where, in the last equality, we have used statement (4) of 
Proposition~\ref{prop:1}. Now, by statement (5) of the same
proposition, we can write
\begin{equation} \label{eq:jacob}
M\qq{0}{0}{\bar{0}} + M\qq{0}{\bar{0}}{0} + M\qq{\bar{0}}{0}{0} \;=\; 0
\end{equation}
(recall that $M\q{0}{\bar{0}} = M\q{\bar{0}}{0} = M\q{0}{0} =
0$). Once again, by statement (4) of 
Proposition~\ref{prop:1}, $M\qq{0}{0}{\bar{0}} =
M\qq{\bar{0}}{0}{0}$. Thus, (\ref{eq:jacob}) reads
$
2M\qq{0}{0}{\bar{0}} + M\qq{0}{\bar{0}}{0} = 0 
$.
Taking the complex conjugate, yields
$$
2\overline{M\qq{0}{0}{\bar{0}}} + M\qq{\bar{0}}{0}{\bar{0}} = 0. 
$$
Finally, using (\ref{eq:10}) and (\ref{eq:1b0}), we get
$M\q{1}{\bar{0}} = 2 \overline{M\q{0}{1}}$. \QED

We have just proven that $M(\calI_4) = 2 \left( \kappa_1^2 \, M\q{0}{1}
  - \overline{M\q{0}{1}}\right)$.  Now we impose $M(\calI_4) = 0$.  Of
course, this will be the case if $M\q{0}{1}=0$ but, in the situation
where $M\q{0}{1}\neq 0$ this can be achieved by fixing $\kappa_1$ as
(see (\ref{eq:I4}))
\begin{equation} \label{eq:biquad} 
\kappa_1 \; = \;
\left(\frac{\overline{M\q{0}{1}}}{M\q{0}{1}}\right)^{1/2} 
\; = \; 
\left(\frac{\overline{M(\calQ_3)}}{M(\calQ_3)}\right)^{1/2} .
\end{equation}
Thus, $\kappa_1$ is a phase: $|\kappa_1| = 1$.  It will be henceforth
assumed that $M\q{0}{1}\neq 0$. If $M\q{0}{1}=0$, $\kappa_1$ has to be
fixed by $M(\calI_5)=0$. We shall not treat this more restrictive case
here.

So far, we have verified the absence of secular terms in the series
expansion (\ref{eq:ansatz}) for $g$ up to order three in $\eps$ and
we have eliminated them from $v_4$ by making a especial choice for the
value of $\kappa_1$ (given by (\ref{eq:biquad})). At this point we
would like to proceed recursively by imposing $M(\calI_n) = 0$, for
all $n \geq 5$. This would give the correct values for the constants
$\kappa_{p}$, $p \geq 2$, and guarantee the absence of secular terms
in all $v_n$, $n \geq 5$. This recursive procedure was used
in~\cite{qp} to eliminate the secular terms from $g$ in cases (I)
and (II). Here, we still have to determine the constants $\kappa_2$
and $\kappa_3$ explicitly (not recursively) before running the
recursive procedure.

\subsection{The Absence of Secular Terms in $v_5$. Fixing $\kappa_2$}

Let us begin by calculating the integrand $\calI_5$ which appears in
$v_5$. Taking $n = 5$ in (\ref{eq:vn}) we get $\calI_5 = 2v_1v_4 +
2v_2v_3$. We first evaluate $v_1v_4$ explicitly and then $v_2v_3$.

For $v_1v_4$, a lengthy computation (see Appendix
\ref{ap:v1v4}) shows that
\begin{eqnarray} \label{eq:v1v4aux2}
v_1 v_4 &=&  
- 6\kappa_1^3\kappa_2 \, \ren{2}\q{0}{2}
+ 2\kappa_1\kappa_2 \,\ren{2}\q{0}{1}
+ i\kappa_1(2\kappa_1 \kappa_3 + \kappa_2^2) \, \calQ_2 
\nonumber \\ & &
+ \kappa_1 \kappa_4 \calQ_0 
+i\kappa_1\calA_1 \, ,
\end{eqnarray}
where $\calA_1$ is the quasi-periodic function defined in
(\ref{eq:defQ15}) and depends only on the constant $\kappa_1$.
Since we are working under the condition
$M(\calQ_0)=M(\calQ_1)=M(\calQ_2)=0$, we can drop the symbol
$\ren{2}$ above.
From this fact and from relation (\ref{eq:M1}), we conclude that 
$M(v_1 v_4)= 2\kappa_1\kappa_2 \, M\q{0}{1} + i\kappa_1M(\calA_1)$.

Let us now calculate the second term in $\calI_5$, namely,
$v_2v_3$. Another lengthy computation (see Appendix
\ref{ap:v2v3}) gives
\begin{eqnarray}
v_2v_3 & = & 
-2i\kappa_1^5 \ren{2}\q{2}{2}  
+2i\kappa_1^3 \ren{2}\q{2}{1}  
-2\kappa_1^3\kappa_2 \ren{2}\q{2}{0}  
+2i\kappa_1^3\ren{2}\q{1}{2}  
\nonumber \\
& &
-2i\kappa_1\ren{2}\q{1}{1}  
+2\kappa_1\kappa_2\ren{2}\q{1}{0}  
-2\kappa_1^3\kappa_2 \ren{2}\q{0}{2}  
\nonumber \\
& &
+2\kappa_1 \kappa_2 \ren{2}\q{0}{1} 
+i(2\kappa_1\kappa_2^2 + \kappa_1^2\kappa_3)\calQ_2 
-i \kappa_3\calQ_1 
+\kappa_2 \kappa_3 \calQ_0 \, .
\label{eq:v2v3aux}
\end{eqnarray}
Clearly the right hand side of equation (\ref{eq:v2v3aux}) is
quasi-periodic.  Since we are working under the condition
$M(\calQ_0)=M(\calQ_1)=M(\calQ_2)=0$, we can drop the symbol $\ren{2}$
above and reorder (\ref{eq:v2v3aux}) in the form
\begin{eqnarray}
v_2v_3 & = & 
 i(2\kappa_1\kappa_2^2 + \kappa_1^2\kappa_3)\calQ_2 
-i \kappa_3\calQ_1 +
\kappa_2 \kappa_3 \calQ_0 
-2i\kappa_1^5 \q{2}{2}  
-2i\kappa_1\q{1}{1}  
\nonumber \\
& &
+2i\kappa_1^3 \left[ \q{2}{1} +   \q{1}{2} \right]
-2\kappa_1^3\kappa_2 \left[ \q{2}{0} + \q{0}{2} \right]  
\nonumber \\
& &
+2\kappa_1\kappa_2 \left[ \q{1}{0} + \q{0}{1} \right]
\, .
\label{eq:Ljkmupv}
\end{eqnarray}
Above we also used $\ren{2}\q{0}{0} =\calQ_2$.  From
(\ref{eq:Ljkmupv}), from the fact that
$M(\calQ_0)=M(\calQ_1)=M(\calQ_2)=0$ and from statements (2) and (3)
of Proposition~\ref{prop:1}, we see immediately that $M(v_2v_3)=0$.

We are now ready to find the value of $\kappa_2$ by imposing the
condition $M(\calI_5)=0$, which guarantees the absence of secular terms
in $v_5$. Since $M(\calI_5)=2M(v_1v_4) + 2 M(v_2v_3)=
4\kappa_1\kappa_2 \, M\q{0}{1} + 2i\kappa_1M(\calA_1)$, we conclude
that
\begin{equation} \label{eq:kappa2}
\kappa_2  \; = \;  - \frac{i M(\calA_1)}{2M\q{0}{1}} \, .
\end{equation}
Note that the right-hand side of (\ref{eq:kappa2}) depends on the
previously fixed $\kappa_1$.

\subsection{The Absence of Secular Terms in $v_6$. Fixing
$\kappa_3$}

We still have to find $\kappa_3$ in order to fix recursively all
$\kappa_n$'s for $n \geq 4$. $\kappa_3$ will be fixed by eliminating
the secular terms from $v_6$, that is, by imposing $M(\calI_6) = 0$.
First of all, we need to write $\calI_6$.  Using relation
(\ref{eq:vn}) for $n = 6$ we find that $\calI_6 = 2v_1v_5 + 2v_2v_4 +
v_3^2$. Let us calculate $2v_1v_5$. Another lengthy computation (see
Appendix \ref{ap:2v1v5}) gives
\begin{eqnarray} \label{eq:v1v5aux}
2v_1v_5 & = & 
\calA_2
-12\kappa_1^3\kappa_3  \ren{2}\q{0}{2}
+ 4\kappa_1\kappa_3\ren{2}\q{0}{1}
\nonumber \\ & &
+4i\kappa_1( \kappa_2 \kappa_3 + \kappa_1 \kappa_4) \calQ_2
+2\kappa_1\kappa_5 \calQ_0 \,,
\end{eqnarray}
where $\calA_2$ is the quasi-periodic function defined in
(\ref{eq:defQ22}) and depends only on the constants $\kappa_1$ and
$\kappa_2$. Since $2v_1v_5$ given above is a sum of quasi-periodic
functions we can take the mean value of each individual term which
appears in the right hand side of (\ref{eq:v1v5aux}) and write
\begin{equation} \label{eq:M2v1v5}
2M(v_1v_5)  \; = \;  M(\calA_2) + 4 \kappa_3 \kappa_1
M\q{0}{1} \, ,
\end{equation}
where, once again, we have used $M(\calQ_0) = M(\calQ_1) = M(\calQ_2)
= 0$ and the identity (\ref{eq:M1}).

We will now calculate the second term in $\calI_5$, namely, $2v_2v_4$.
We have (see Appendix \ref{ap:2v2v4})
\begin{eqnarray}  \label{eq:2v2v4aux}
2 v_2 v_4 &=&
\calA_3
-4  \kappa_1^3 \kappa_3  \ren{2}\q{2}{0}
+4 \kappa_1   \kappa_3  \ren{2}\q{1}{0}
\nonumber \\
& &
+ 2i( \kappa_1^2 \kappa_4 +2 \kappa_1   \kappa_2 \kappa_3 ) \calQ_2 
  - 2i\kappa_4\calQ_1 
  +  2\kappa_2\kappa_4 \calQ_0 
 \, ,
\end{eqnarray}
where $\calA_3$ is defined in (\ref{eq:defQ26}) and depends only
on the constants $\kappa_1$ and $\kappa_2$.

We can now proceed and take the mean value of $2v_2v_4$ from
(\ref{eq:2v2v4aux}). Using $M(\calQ_0) = M(\calQ_1) =
M(\calQ_2) = 0$ and (\ref{eq:M1}), we
get
\begin{equation} \label{eq:M2v2v4}
2M(v_2v_4) \; = \;  M(\calA_3) + 4 \kappa_1 \kappa_3 M\q{0}{1}  \, .
\end{equation}

We are almost through with the calculation of $\calI_6$.
We still need $v_3^2$. Using relations 
(\ref{eq:v1ren})--(\ref{eq:2v1v2}),
\begin{eqnarray} 
v_3 & = &
q^{-1}\left\{ 2i \ren{2}\q{0}{v_1v_{2}}+ \kappa_3 \calQ_0 \right\}
\nonumber \\
&= &
q^{-1}
\left\{ 
-2\kappa_1^3\ren{2}\q{0}{2}
+2\kappa_1\ren{2}\q{0}{1}
+2i \kappa_1\kappa_2 \calQ_2
+ \kappa_3 \calQ_0
\right\}.
\label{eq:v3especial}
\end{eqnarray}
Hence,
\begin{eqnarray} 
v_3^2 & = &
\kappa_3^2 \calQ_0
+ 4\kappa_1\kappa_3 
\left[
-\kappa_1^2\ren{2}\q{0}{2}
+\ren{2}\q{0}{1}
+i\kappa_2 \calQ_2
\right]
- 4\calA_4 \, ,
\nonumber
\end{eqnarray}
where
\begin{equation} 
\calA_4 \; := \; 
\calQ_0^{-1}
\left[
i\kappa_1^3\ren{2}\q{0}{2}
-i\kappa_1\ren{2}\q{0}{1}
+ \kappa_1\kappa_2 \calQ_2
\right]^2 \, .
\nonumber
\end{equation}
Again, $\calA_4$ is quasi-periodic and depends only on the
already fixed constants $\kappa_1$ and $\kappa_2$. Therefore,
using once more $M(\calQ_0) = M(\calQ_2) = 0$ and (\ref{eq:M1}), we get
\begin{equation} \label{eq:Mv32}
M(v_3^2)  \; = \;  4 \kappa_3 \kappa_1 M\q{0}{1} - 4 M(\calA_4) \, .
\end{equation}

Finally, imposing
$M(\calI_6)=0$, i.e., $2M(v_1v_5) + 2M(v_2v_4) + M(v_3^2) = 0$,
we obtain from our previous calculations (relations (\ref{eq:M2v1v5}),
(\ref{eq:M2v2v4}) and (\ref{eq:Mv32})),
\begin{equation} \label{eq:kappa3}
\kappa_3  \; = \;  \frac{4M(\calA_4) - M(\calA_3) -
M(\calA_2)} {4\kappa_1 M\q{0}{1}} \, .
\end{equation}
Note that the right-hand side of (\ref{eq:kappa3}) depends on the
previously fixed $\kappa_1$ and $\kappa_2$.

\subsection{The Absence of Secular Terms in $v_n$, $n \geq
7$. Fixing $\kappa_{n-3}$ Recursively}

So far, we have fixed the constants $\kappa_1$, $\kappa_2$ and
$\kappa_3$ individually. Now we proceed to fix recursively all other
$\kappa_{n-3}$ for all $n \geq 7$. We have to impose
\begin{equation} \label{eq:condicao}
M(\calI_n)  \; = \;  0 \quad \Longrightarrow \quad
M\left(\sum_{p=1}^{n-1} v_p v_{n-p}\right)  \; = \;  0 \, 
\end{equation}
for all $n \geq 7$. Condition (\ref{eq:condicao}) guarantees the
absence of secular terms in all $v_n$, $n \geq 7$.

The idea now is to use (\ref{eq:condicao}) to calculate recursively
the constants $\kappa_{n-3}$, for all $n \geq 7$, that is, $\kappa_4,
\; \kappa_5, \; \ldots$. Of course, we already have $\kappa_1$,
$\kappa_2$ and $\kappa_3$ and, hence, we completely know $v_1$, $v_2$
and $v_3$.  We also know that all functions, from $v_1$ to $v_6$, are
quasi-periodic. From now on we will work inductively. Thus, it will be
supposed for each $n \geq 7$ that we fixed $\kappa_1, \; \ldots , \;
\kappa_{n-4}$ by imposing $M(\calI_m)=0$ for all $m=2, \, \ldots , \,
n-1$ and that, as a consequence, all functions $v_1, \; \ldots , \;
v_{n-1}$ are quasi-periodic. Note that it is indeed true for $n = 7$.

By our inductive hypothesis, we are allowed to take the summation
out of the mean value $M$ in (\ref{eq:condicao}) and write
\begin{equation} \label{eq:somaaberta}
2\left[ M(v_1 v_{n-1}) +  M(v_2 v_{n-2}) +  M(v_3 v_{n-3})\right] +
\sum_{p=4}^{n-4} M(v_p v_{n-p})  \; = \;  0 \, ,
\end{equation}
where, by convention, ${\sum_{p=4}^{n-4} M(v_p
v_{n-p}) = 0}$ for $n = 7$. Let us introduce now the following
definition:
\begin{equation} \label{eq:ln}
l_m(t) \defi q(t)(v_m(t) - \kappa_m q(t)) \, ,
\end{equation}
for all $m \geq 4$. Note that, by relation (\ref{eq:vn}), the
functions $l_m$'s above can also be written as
\begin{equation} \label{eq:lntab}
l_m(t)  \; = \;   i \, q(t)^2 \left(\int_0^t \sum_{p=1}^{m-1}
v_p(\tau) v_{m-p}(\tau) \, d\tau \right) \, .
\end{equation}
For $m<n$ we are allowed to write
\begin{eqnarray} \label{eq:lntab2}
l_m(t)  &=&   i \, q(t)^2 \left(\int_0^t \sum_{p=1}^{m-1}
\left[v_p(\tau) v_{m-p}(\tau) -M\left(v_pv_{m-p}\right)\right] \, d\tau \right)
\nonumber \\ 
&=& i \sum_{p=1}^{m-1}\ren{2}\q{0}{v_pv_{m-p}}_t ,
\end{eqnarray}
since we assumed $M(\sum_{p=1}^{m-1}v_pv_{m-p})=M(\calI_m)=0$ for all 
$m<n$, by the inductive hypothesis. Hence, $l_m$ are quasi-periodic for
all $m<n$, by the inductive hypothesis.

Let us use the definition given in (\ref{eq:ln}) and evaluate the
first three terms which appear in (\ref{eq:somaaberta}). Beginning
with the first one, we have
\begin{equation} 
M(v_1 v_{n-1}) \;=\; M\left(v_1(q^{-1}l_{n-1} + \kappa_{n-1}q)\right)
\; = \; \kappa_1 M( l_{n-1} ) \, ,
\label{eq:somando1pm}
\end{equation}
where we have used (\ref{eq:v1}) and the fact that $M(\calQ_0) = 0$.
Using (\ref{eq:ln}), the second term of (\ref{eq:somaaberta}) can
be evaluated as
\begin{equation} 
M(v_2 v_{n-2}) \;=\; M\left(v_2(q^{-1}l_{n-2} + \kappa_{n-2}q)\right)
\; = \; 
M(q^{-1}v_2 l_{n-2}) \, ,
\label{eq:somando2pm}
\end{equation}
where we have used (\ref{eq:v2ren}) to express $qv_2$ and the fact that
$M(\calQ_0) = M(\calQ_1) = M(\calQ_2) = 0$. 
Finally, for the third term of (\ref{eq:somaaberta}), we have
\begin{equation} 
M(v_3 v_{n-3}) \;=\; M\left(v_3(q^{-1}l_{n-3} + \kappa_{n-3}q)\right)
\; =\; M(q^{-1}v_3 l_{n-3}) + \kappa_{n-3} M(qv_3 ) \, .
\label{eq:somando3}
\end{equation}
The product $qv_3$ can be obtained from (\ref{eq:v3especial}), from
which we conclude that $ M(qv_3) = 2\kappa_1M\q{0}{1}$. Inserting this
into (\ref{eq:somando3}), gives
\begin{equation} \label{eq:somando3pm}
M(v_3 v_{n-3})  \; = \;  M(q^{-1}v_3 l_{n-3}) + 2 \kappa_1 \kappa_{n-3} 
M\q{0}{1} \, .
\end{equation}

The substitution of (\ref{eq:somando1pm}), (\ref{eq:somando2pm})
and (\ref{eq:somando3pm}) into (\ref{eq:somaaberta}), gives us
\begin{multline} \label{eq:analisechave}
2 \, \left\{ \, \underbrace{\kappa_1M(l_{n-1})}_{\text{(i)}} +
\underbrace{M(q^{-1}v_2 l_{n-2})}_{\text{(ii)}} +
\underbrace{M(q^{-1}v_3 l_{n-3})}_{\text{(iii)}}  +
\underbrace{2 \kappa_{n-3} \kappa_1 M\q{0}{1}}_{\text{(iv)}} \,
\right\} 
\\ 
+ \; \underbrace{\sum_{p=4}^{n-4} M(v_p v_{n-p})}_{\text{(v)}}  \; =
\; 0 \, . 
\end{multline}

Before we proceed, let us make some comments on our strategy. Equation
(\ref{eq:analisechave}) is a direct consequence of (\ref{eq:condicao})
and, thus, is being imposed for each $n \geq 7$, leading to the values
of $\kappa_4$, $\kappa_5$ and so on. By our induction hypothesis, we
have fixed the constants $\kappa_1, \ldots, \kappa_{n-4}$ and, hence,
we completely know $v_1, \ldots, v_{n-4}$.  For this reason the terms
{\rm (iii)} and {\rm (v)} are known by assumption (by
(\ref{eq:lntab2}), the evaluation of $l_{n-3}$ requires the knowledge
of $v_1, \ldots, v_{n-4}$).  Our aim is to use (\ref{eq:analisechave})
as a condition to fix $\kappa_{n-3}$ and we, therefore, have to
isolate the dependence of (\ref{eq:analisechave}) on $\kappa_{n-3}$.
The function $l_{n-2}$, in term {\rm (ii)}, depends implicitly on
$v_{n-3}$ and, hence, on $\kappa_{n-3}$ (see, again, relation
(\ref{eq:lntab2})). The term {\rm (i)}, however, depends implicitly on
$\kappa_{n-2}$ and $\kappa_{n-3}$. This dependence on $\kappa_{n-2}$
could be a problem, since we are still working to fix $\kappa_{n-3}$.
Nevertheless, as will be shown, the conditions $M(\calQ_1) = 0$ and
$M(\calQ_2) = 0$ fortunately eliminate $\kappa_{n-2}$ from the final
expression, and we will be led to a condition expressing
$\kappa_{n-3}$ in terms of known quantities.

Let us now compute terms  {\rm (i)} and  {\rm (ii)}.
After a long computation, found in Appendix \ref{ap:somando1}, we
get
\begin{equation} \label{eq:somando1}
\kappa_1M( l_{n-1})  \; = \;  2\kappa_1 \kappa_{n-3} M\q{0}{1} +
\calR^{(1)}_n \, ,
\end{equation}
where $\calR^{(1)}_n$ is defined in (\ref{eq:R1def}) and depends
on constants $\kappa_1, \, \ldots, \, \kappa_{n-4}$ only.  For term
{\rm (ii)} of (\ref{eq:analisechave}) we get, after another long
computation presented in Appendix \ref{ap:somando2},
\begin{equation} \label{eq:somando2}
M(q^{-1} v_2 l_{n-2}) \; = \; -2 \kappa_1 \kappa_{n-3} M\q{0}{1} 
+ \calR^{(2)}_n  \, ,
\end{equation}
where $\calR^{(2)}_n $ is defined in (\ref{eq:R2def}). We again stress
that $\calR^{(2)}_n$ depends on $\kappa_1, \, \ldots, \, \kappa_{n-4}$
only.

We are now ready to give the precise value of $\kappa_{n-3}$ in order
to satisfy (\ref{eq:condicao}). Collecting (\ref{eq:somando1})
and (\ref{eq:somando2}) and inserting them into
(\ref{eq:analisechave}), we obtain
\begin{equation} 
\kappa_{n-3} \; = \; 
\frac{-1}{4\kappa_1 M\q{0}{1} }
\left\{      
\sum_{p=4}^{n-4}M(v_p v_{n-p}) + 2\calR^{(1)}_n +
       2\calR^{(2)}_n + 2M(q^{-1}v_3 l_{n-3})
\right\}
\, , \nonumber
\end{equation}
for all $n \geq 7$. Note that $\calR^{(1)}_n$ and $\calR^{(2)}_n$ can
be completely computed in any order of recursion (see equations
(\ref{eq:R1def}) and (\ref{eq:R2def})). The function $l_{n-3}$ can
also be computed in any order of recursion if one uses relation
(\ref{eq:lntab2}).

Summarising our conclusions, under conditions ($\mbox{III}_0$) and
$M\q{0}{1}\neq 0$, i.e., under condition (III) of
Theorem~\ref{teo:condIII}, and with the constants $\kappa_n$
recursively chosen as
\begin{eqnarray}
\kappa_1 & = &
\left(\frac{\overline{M\q{0}{1}}}{M\q{0}{1}}\right)^{1/2}, 
\label{eq:capa123}
\\
\kappa_2 &=& - \frac{i M(\calA_1)}{2M\q{0}{1}} \,\, , \qquad
\kappa_3 \;=\; \frac{4M(\calA_4) - M(\calA_3) -
M(\calA_2)} {4\kappa_1 M\q{0}{1}} \, ,  
\\
\kappa_{n-3} &=& 
\frac{-1}{4\kappa_1 M\q{0}{1} }
\left\{      
\sum_{p=4}^{n-4}M\left(v_p v_{n-p}\right) + 2\calR^{(1)}_n +
       2\calR^{(2)}_n + 2M\left(q^{-1}v_3 l_{n-3}\right)
\right\}, \nonumber
\\ 
\label{eq:capan}
\end{eqnarray}
for $n \geq 7$, all secular terms are eliminated from the formal
solution (\ref{eq:ansatz}) of the generalised Riccati equation
(\ref{eq:riccati}). Notice that hypothesis $M\q{0}{1}\neq 0$ is the
\underline{only} additional restriction needed to
(\ref{eq:capa123})-(\ref{eq:capan}). The proof of
Theorem~\ref{teo:condIII} is thus complete. $\QED$


\section{The Case of the Monochromatic Field}
\label{sec:example} 
\zerarcounters

We illustrate our method and our results considering the simplest case
of monochromatic interactions (ac-dc field)
\begin{equation} \label{eq:interaction}
  f(t) \; = \; F_0 + \varphi \cos(\omega t) ,
\end{equation}
which are of particular importance in physical applications.  We want
to show that with conditions (I)-(III) we obtain with our method
convergent perturbative solutions of this problems for all parameters
$F_0$ and $\varphi$, except perhaps for some spurious situations.
For (\ref{eq:interaction}) one has
\begin{equation*}
\calQ_0 (t) \; =\; \sum_{n \in \Z}J_n\left(\chi_1\right)
e^{i(n + \chi_2)\omega t}\, ,
\end{equation*}
where $J_n$ is the Bessel function of first kind and order $n$ and
where we defined $\chi_1 := 2\varphi/\omega$, $\chi_2 := 2F_0/\omega$.
Hence, condition (I) (treated in detail in \cite{pp}) holds provided
$\chi_2 = -m$, with $m$ integer, and provided $\chi_1$ is not a zero
of the Bessel function $J_m$.  Here, we have by
(\ref{eq:Omegacasogeral}), 
$$
\Omega \; = \; -\frac{m\omega}{2} + \eps J_{m}(\chi_1) + O(\eps^2)\, .
$$
See also the discussion in \cite{Sacchetti}.  Let us consider the
complementary situations.

{\bf i.} Consider the case where $\chi_2=-m$, a non-zero integer, and
$\chi_1$ is a zero of $J_m$.  One has $M(\calQ_0)=0$ and we have to
look first at $M(\calQ_1)$.  We get,
$$
M(\calQ_1)\; =\; \sum_{k\in\Z_\ast}\frac{J_{k+m}(\chi_1)^2}{ik\omega}\, .
$$
For integer $m$ one has
\begin{equation}
\sum_{k\neq 0}\frac{J_{k+m}(x)^2}{k} \; = \;
J_m(x)
\left[ 
\left. -2\frac{\partial}{\partial \nu}J_{\nu}(x)\right|_{\nu=m} +
\pi Y_m(x)
\right],
\label{eq:identidadeBessel}
\end{equation}
where 
$\pi Y_m(x) = \left. \left( \frac{\partial}{\partial \nu}J_{\nu}(x)
    -(-1)^m\frac{\partial}{\partial \nu}J_{-\nu}(x)
  \right)\right|_{\nu=m}$, 
$Y_m$ being Bessel functions of second kind.  
Identity (\ref{eq:identidadeBessel}) is apparently unmentioned in the
literature, and we sketch its proof in Appendix \ref{identidadeBessel}.
Since $\chi_1$ is bound to be a zero of
$J_m$, one concludes from this identity that $M(\calQ_1)=0$ in this
case. A direct computation shows that 
$$
M\q{0}{1}  =  \frac{1}{\omega^2} \calT_m(\chi_1),
\;\;\mbox{ where }\;\;
\calT_m(x) := - \sum_{k, p \neq 0} \frac{ J_{m+p}(x) J_{m+p-k}(x) J_{m+k}(x)
  }{ kp }.
$$
Numerical calculations indicate that $\calT_m(x)$ does not vanish
at the zeros of $J_m$.  We refrain from giving an analytic proof of
this fact.  We conclude that condition (III) holds in case {\bf i},
except, perhaps, for spurious zeros of $J_m$ for which $\calT_m$
eventually vanishes, and whose existence could not be ruled out
numerically.  By (\ref{eq:OmegacasoII}),
$$
\Omega \; = \; -\frac{m\omega}{2} +
\frac{2\eps^3}{\omega^2}\calT_m(\chi_1)
+O(\epsilon^4)\, .
$$

{\bf ii.} Consider the case where $\chi_2$ is non-integer
(see also the discussion in \cite{Sacchetti}). One has
$M(\calQ_0)=0$ and we have to look at $M(\calQ_1)$.  We get,
$$
M(\calQ_1)\; = \; \frac{i}{\omega\chi_2} \left[ J_0(\chi_1)^2 +
  2\chi_2^2\sum_{k=1}^\infty\frac{J_k(\chi_1)^2}{\chi_2^2 - k^2}
\right]\, .
$$
Generally the r.h.s is non-zero and we have condition (II). Hence, by
(\ref{eq:OmegacasoII}),
\begin{equation}\label{eq:Omegaii}
\Omega \; = \; F_0 + \frac{\eps^2}{\omega\chi_2} \left[ J_0(\chi_1)^2 +
  2\chi_2^2\sum_{k=1}^\infty\frac{J_k(\chi_1)^2}{\chi_2^2 - k^2}
\right] + O(\eps^3) \, .
\end{equation}

Note, however, that on each interval $\chi_2 \in (k, \; k+1)$, $k=1,
\, 2, \ldots$, the terms $ \frac{J_k(\chi_1)^2}{\chi_2^2 - k^2}
+\frac{J_{k+1}(\chi_1)^2}{\chi_2^2 - (k+1)^2} $ vary continuously from
$+\infty$ to $-\infty$. Hence, there is on each interval $(k, \;
k+1)$, $k=0, \, 1,\, 2, \ldots$, a special value $\chi_2^s$ of
$\chi_2$ (depending on $\chi_1$) for which $M(\calQ_1)=0$, and we
would be out of case (II). But when $2\chi_2$ is non-integer, one has
$M(\calQ_3)=0$. Hence, except for the very unlikely case where
$2\chi_2^s$ is an integer, we would be out of condition (III) as well,
and $\Omega =F_0 + O(\epsilon^4)$. This special case may not be very
interesting, because $\chi_2$ has to be chosen with precision. That
could be difficult to fix it in some experimental setting.

Another special situation would occur when $\chi_2$ is chosen to
satisfy $F_0 - i \eps^2M(\calQ_1)=0$. By the argument above, this is
possible, but $\chi_2$ will depend on $\eps$. It is therefore unclear
if $\Omega$ will be just $O(\eps^4)$ or ``small'' (eventually leading
to an even stronger dynamical localisation than we have in case
(III)). It is not even clear that we will be in a situation where our
series converge, and we left this other special situation without more
comments.

It is interesting to compare the expressions for the secular frequency
$\Omega$ in the three situations above (for $F_0\neq 0$) with the
situation where $\varphi=0$, where the secular frequency $\Omega_0$ is
$\Omega_0 := F_0\sqrt{1+(\frac{\eps}{F_0})^2} = F_0 +
\frac{\eps^2}{2F_0}+O(\eps^3)$. This reveals the effect of the
ac-field $\varphi\cos(\omega t)$ on the secular frequency.
Taking $\chi_1\to0$ in (\ref{eq:Omegaii}) we recover $\Omega_0$.

\begin{figure}[h!]
\centering{
\epsfig{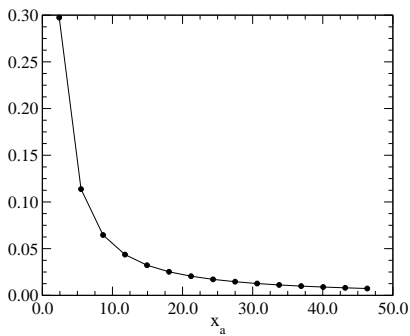}
\caption{The filled circles represent the values of 
  $\left|\calT\left(x_a\right)\right|$ as a function of the $a$-th
  zero $x_a$ of $J_0$ in $\R_+$. We infer an exponential decay of
  $|M\q{0}{1}|$ as $a$ increases.}
\label{fig:conditionIII}
}
\end{figure}

{\bf iii.} Consider the case where $\chi_2=0$, i.e., $F_0=0$, and
$\chi_1=x_a$, the $a$-th zero zero of the Bessel function $J_0$ on the
positive real axis. This case is interesting in connection with the
issue of dynamical localisation, as discussed by many authors
(\cite{Grossmann}. For more references, see
\cite{BarataWreszinski2,Sacchetti}).  Here $M(\calQ_0)=0$ and,
\begin{equation*}
  M(\calQ_1) \;= \; \frac{i}{\omega} \sum_{m = 1}^{\infty}
  \left(\frac{J_m(\chi_1)^2 - J_{-m}(\chi_1)^2}{m}\right) \; = \; 0\, ,
\end{equation*}
since $J_k(x)=(-1)^kJ_{-k}(x)$.  Thus, condition (II) does not apply
and we have to look at $M\q{0}{1}$. We obtain $M\q{0}{1} =
\frac{1}{\omega^2} \calT\left(x_a \right)$, with
$$
\calT\left(x_a \right) \; := \; -\sum_{n, m \in \Z_{\ast}}
\frac{J_n\left(x_a \right) J_{n-m}\left(x_a \right) J_{m}\left(x_a
  \right) }{n \, m}\, .
$$
We conclude that condition (III) will be valid, except perhaps for
spurious zeros of $J_0$ for which $\calT\left(x_a \right)$ eventually
vanishes.  Numerical computations, though, indicate that such zeros
may not exist.  We abstain from giving an analytical proof that
$\calT\left(x_a \right) \neq 0$ for all zeros of the $J_0$ function.
Figure~\ref{fig:conditionIII} shows the values of
$\left|\calT\left(x_a \right)\right|$ calculated numerically on the
first fifteen zeros $x_a$ of $J_0$. We note an exponential decay of
$\left|\calT\left(x_a \right)\right|$ as $x_a$ increases, suggesting
that $|\calT\left(x_a \right)| \to 0$ only for $a \to \infty$.

We conclude that condition (III) is suitable for studying the
monochromatic field when $\chi_1$ lies over the ``resonant'' points
$x_a$, leading, together with condition (I), to a complete solution of
(\ref{eq:interaction}) except, perhaps, for some rather spurious
situations.

Note finally that in this case {\bf iii} we have $\Omega
=O(\epsilon^3)$ (see (\ref{eq:OmegacasoIII})).  In fact, the first
contribution to $\Omega$ is $2\frac{\epsilon^3}{\omega^2}
\calT\left(x_a\right)$. This weak dependence on $\epsilon$ implies
long transition times for certain probability amplitudes (see below),
an issue known as dynamical localisation (see \cite{Sacchetti}).

In order to test our algorithm and to extract more information from
our solutions, we wrote a computer program to compute the matrix
elements of the propagator $U(t)$ given in (\ref{U11U12}) for the case
{\bf iii} described above, where one has approximate dynamical
localisation. The results are excellent and are briefly reported in the
next section. 

\begin{remark}
  We obtained for case {\bf iii}, $\Omega
  =2\frac{\epsilon^3}{\omega^2} \calT\left(x_a\right)+O(\eps^4)$.
  Following another path, we obtained in~\cite{BarataWreszinski2},
$$\displaystyle \Omega = -2\frac{\epsilon^3}{\omega^2} \sum_{ n_1, \,
    n_2 \in {\mathbb Z} } \frac{J_{2n_1+1}(x_a) J_{2n_2 +1}(x_a)
    J_{-2(n_1+n_2 +1)}(x_a)}{(2n_1+1)(2n_2+1)]} + O(\eps^4).
$$ Thus, we must have
\begin{multline}
\sum_{n, m \in \Z_{\ast}} \frac{J_n\left(x_a \right)
  J_{n-m}\left(x_a \right) J_{m}\left(x_a \right) }{n \, m}\, \; = \;
\\
\sum_{ n_1, \, n_2 \in {\mathbb Z} } \frac{J_{2n_1+1}(x_a)
  J_{2n_2+1}(x_a) J_{-2(n_1+n_2 +1)}(x_a) }{(2n_1+1)(2n_2+1)}\, .
\nonumber
\end{multline}
Curiously, such an identity is not easy to prove by direct means.
It can be established, however, with the use of statement (6) of
Proposition \ref{prop:1}, namely, from $M\qq{0}{0}{0}=0$.  Actually,
such an identity is not specific for Bessel functions, but is valid for
sequences satisfying certain properties.
\end{remark}


\section{Numerical Results}
\label{sec:numerical} 
\zerarcounters

In this final section, we will briefly show some numerical results
that were obtained using condition (III) to study the monochromatic
interaction over the resonant points (see the discussion in the last
section). Our goal here is to show how one can numerically use the
results of Theorems~\ref{teo:riccati},~\ref{teo:condIII}
and~\ref{teo:convergence} to produce results of physical interest.

The computation of the wave function $\Phi(t)$ associated with the
Schr\"odinger equation (\ref{equacaodeSchroedingerparaPhi}) can be
performed by means of the propagator $U(t)$, given in
(\ref{definicaodeU}), in terms of the solution of the generalised
Riccati equation (\ref{eq:riccati}) given in
Theorem~\ref{teo:condIII}. Due to the uniform convergence of $g(t)$
for all $t \in \R$ (see Theorem~\ref{teo:convergence}), one can use
the proper expressions given in~\cite{pp} to compute the elements of
$U(t)$ via uniformly convergent power series in $\eps$. Although this
task may not be so trivial, it can be implemented numerically with
great success, as we are about to show now.

The first step toward the computation of $U(t)$ is to evaluate the
Fourier coefficients of $g(t)$ using the recursive relations given in
(\ref{PerVFourier1})-(\ref{PerVFouriern}). Of course, we need first to
compute the constants $\kappa_n$. To this end, we have to use
expressions (\ref{eq:capa123}) and (\ref{eq:capan}). Those equations
involves the mean value of some previously defined functions, which
can be obtained trough the zero order terms of their Fourier
decomposition. Equation (\ref{eq:decomp}) of Proposition~\ref{prop:1}
can be used to this end. Once $V_m^{(n)}$ are all computed, the $m$-th
Fourier coefficient of order $n$ of $g(t)$ can be obtained via the
convolution $G_m^{(n)} = \sum_{p \in \Z} Q_{m-p} V_{p}^{(n)}$ (see
equation (\ref{eq:ansatz})). As a second and final step towards the
computation of $U(t)$, we have to use the expressions given
in~\cite{pp}. Basically, these expressions convert the integral form
of the elements of the propagator $U(t)$ (see
Theorem~\ref{teo:riccati}) into convergent power series in $\eps$
using the Fourier coefficients $G_m^{(n)}$ of $g(t)$. They also lead
to the expression of $U(t)$ in terms of its {\it Floquet form} (see
Theorem~\ref{teo:floquet}).  A more detailed description of our
numerical study will be postponed for a future
publication~\cite{dacjcab2}. For now, let us briefly show some of our
results.

We have followed steps one and two above to numerically compute the
propagator $U(t)$ associated with (\ref{equacaodeSchroedingerparaPhi})
when $f(t)=\varphi \cos(\omega t)$ in the situation where $M(\calQ_0)
= 0$, that is, over the ``resonant'' points defined by the condition
$\varphi = \omega x_a /2$, $x_a$ being the $a$-th positive zero of
$J_0$.

Let $\Phi_+ = \left( { 1 \atop 0} \right)$ and $\Phi_- = \left( { 0
    \atop 1} \right)$ be two orthogonal states of a system described
by (\ref{equacaodeSchroedingerparaPhi}).  The probability for the
transition from the initial state $\Phi_+$ to the final state
$\Phi_-$ at time $t$ is $P(t) := \left|\,\langle \Phi_+, \, U(t)
  \Phi_-\rangle\,\right|^2 = \left|U_{12}(t)\right|^2$. We computed
$P(t)$ numerically using our expansions. To estimate the accuracy of
our calculations, we tested the unitarity of the time evolution and
considered the quantity $ N(t) := |U_{11}(t)|^2 + |U_{12}(t)|^2 $, which
should be identically equal to $1$ for unitary $U(t)$.

\begin{figure}[h!]
\centering{
\epsfig{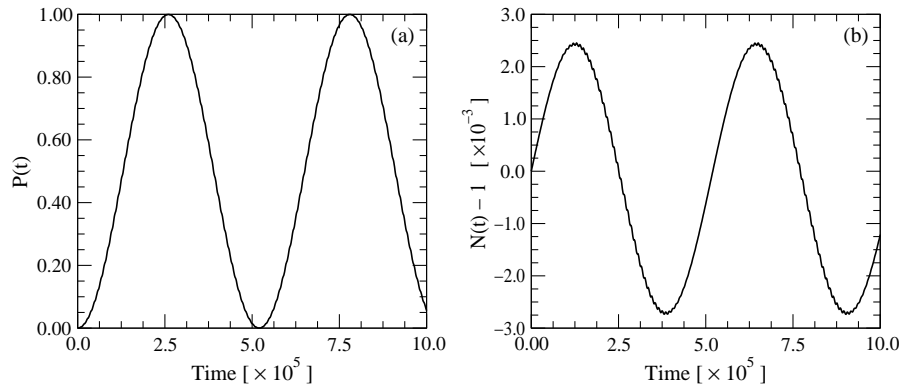}
\caption{{\bf (a)} The transition probability $P(t)$ as function of
  time. Here $\eps = 0.1$, $\omega = 10$ and $2\varphi_1/\omega =x_1$,
  the first zero of $J_0$.  {\bf (b)} The quantity $N(t)-1$
  that measures deviation from unitarity.}
\label{fig:P1}
}
\end{figure}

In Figure~\ref{fig:P1}{\bf(a)}, we show $P(t)$ for $\varphi = \omega
x_1 / 2$, $x_1$ being the first positive zero of $J_0$. We took
$\omega = 10.0$, $\eps = 0.1$ and worked with a sixth order expansion
in $\eps$. The time interval considered corresponds to $1.6 \times
10^6$ times the basic cycle $2\pi/\omega$ of $f$. We recall that in
this case we have condition (III) and the {\it secular frequency},
which dominates the quasi-periodic evolution of the system, is of
order $\eps^{3}/\omega^2$ (see \cite{BarataWreszinski2} or the
discussion in Section \ref{sec:secfreq}). This explains the long time
($\sim 10^6$ of the basic cycle of $f$) needed for the system to
transit from $\Phi_+$ to $\Phi_-$.

In Figure~\ref{fig:P1}{\bf(b)}, we show the quantity $N(t)-1$ as a
function of time obtained from the same calculations leading to
Figure~\ref{fig:P1}{\bf(a)}. By looking at the deviations of $N(t)$
from $1$, we can infer that our perturbative solution produces errors
of the order of only $0.3 \%$. This excellent numerical precision for
long times is a consequence of the uniform convergence in time of our
expansions, i.e., of the elimination of the secular terms.

\begin{figure}[h!]
\centering{
\epsfig{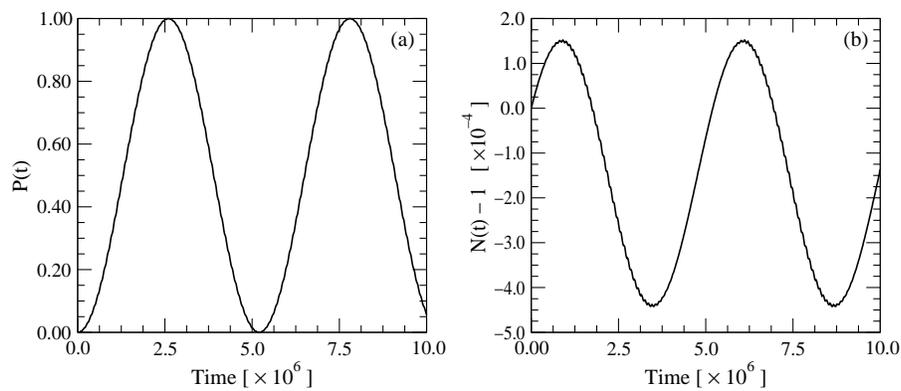}
\caption{{\bf (a)} The transition probability $P(t)$ as function of
  time. Here $\eps = 0.01$, $\omega = 1.0$ and $2\varphi_1/\omega =x_1$,
  the first zero of $J_0$.  {\bf (b)} The quantity $N(t)-1$
  that measures deviation from unitarity.}
\label{fig:P2}
}
\end{figure}

\begin{figure}[h!]
\centering{
\epsfig{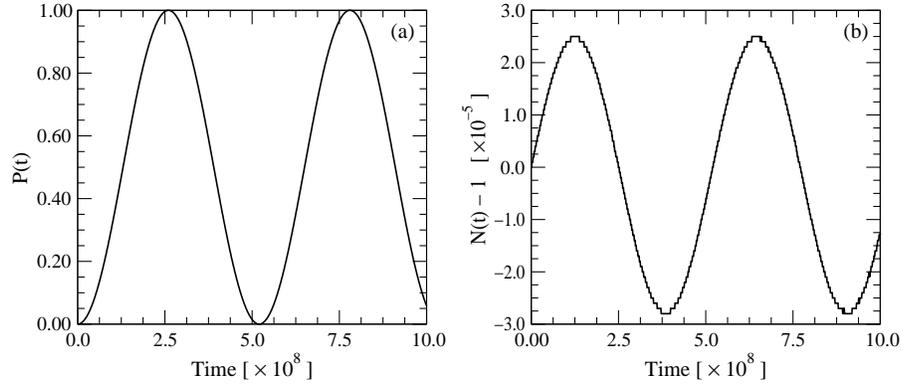}
\caption{{\bf (a)} The transition probability $P(t)$ as function of
  time. Here $\eps = 0.01$, $\omega = 10.0$ and $2\varphi_1/\omega =x_1$,
  the first zero of $J_0$.  {\bf (b)} The quantity $N(t)-1$
  that measures deviation from unitarity.}
\label{fig:P3}
}
\end{figure}

In Figure~\ref{fig:P2} we took $\omega = 1.0$, $\eps = 0.01$ and
$\varphi_1 = \omega x_1 / 2$. From Figure~\ref{fig:P2}{\bf(b)} we
infer errors of the order of only $0.05 \%$. The time interval
considered corresponds to $1.6\times 10^6$ times the basic cycle
$2\pi/\omega$ of $f$.
In Figure~\ref{fig:P3} we took $\omega = 10.0$, $\eps = 0.01$ and
again $\varphi_1 = \omega x_1 / 2$. From Figure~\ref{fig:P3}{\bf(b)} 
we infer errors of the order of only $0.003 \%$. The time interval
considered corresponds to $1.6\times 10^9$ times the basic cycle
$2\pi/\omega$ of $f$.

The numerical computations above involved the evaluation of the Fourier
expansions of several functions, like $q$, $v_n$'s etc. We typically
computed about 20 terms of that expansions, but since the coefficients
decay very fast, even less terms are needed. More details about the
numerical computations will be presented in a future
publication~\cite{dacjcab2}.

\newpage

\vspace{0.5cm}

\begin{appendix}

\begin{Large}
\noindent{\bf Appendices }
\end{Large}

\section{Some Special Relations}
\label{ap:VariasProvas}
\zerarcounters 

This appendix presents the proofs of some relations used in
Section~\ref{sec:case}. Since they involve a somewhat large amount of
algebraic manipulations we prefer to separate them from the main text.
In the following we will (often without explicit mention) make
repeated use of the propositions and corollaries proven in
Sections~\ref{sec:mean} and~\ref{sec:RenOp}.

\subsection{Obtaining Relation (\ref{eq:v1v4aux2})}
\label{ap:v1v4}

Since we have $M(\calI_3)=0$ and since with the choice of $\kappa_1$
in (\ref{eq:biquad}) we imposed $M(\calI_4)=0$, we have by the
recursive relations (\ref{eq:v1ren})-(\ref{eq:vnren})
\begin{eqnarray} 
v_1v_3  & = & 2i\kappa_1 \ren{2}\q{0}{v_1v_2} 
              + \kappa_1 \kappa_3 \calQ_0 \, ,
\label{eq:niiiuHg2}
\\
v_1 v_4 & = &   2i\kappa_1 \ren{2}\q{0}{v_1v_3} 
              + i\kappa_1 \ren{2}\q{0}{v_2^2} 
              + \kappa_1 \kappa_4 \calQ_0 \, .
\label{eq:niiiuHg1}
\end{eqnarray}
Inserting (\ref{eq:2v1v2}) into the r.h.s. of (\ref{eq:niiiuHg2}) we
get
\begin{equation}
v_1v_3  \; = \; 
-2\kappa_1^4 \, \ren{2}\q{0}{2} 
+2\kappa_1^2 \, \ren{2}\q{0}{1}
+2i\kappa_1^2\kappa_2 \, \ren{2}\q{0}{0}
+ \kappa_1 \kappa_3 \, \calQ_0 \, .
\label{eq:v1v3util}
\end{equation}
Inserting this into the r.h.s. of (\ref{eq:niiiuHg1}) gives
\begin{eqnarray}
v_1 v_4 & = & 
  -4i\kappa_1^5 \, \ren{3}\qq{0}{0}{2} 
+ 4i\kappa_1^3 \, \ren{3}\qq{0}{0}{1}
- 4\kappa_1^3\kappa_2 \, \ren{3}\qq{0}{0}{0}
\nonumber
\\
 & & 
     + 2i\kappa_1^2 \kappa_3 \, \ren{2}\q{0}{0}
     + i\kappa_1 \ren{2}\q{0}{v_2^2}
     + \kappa_1 \kappa_4 \calQ_0 \, .
\label{eq:iniowDf1}
\end{eqnarray}
Let us now compute $v_2^2$. 
Since $v_2 = \kappa_2 q + iq^{-1}(\kappa_1^2\calQ_2 - \calQ_1)$, we have
\begin{equation}
v_2^2 \; = \; 
  \kappa_2^2 \calQ_0 
+ 2i\kappa_1^2\kappa_2\calQ_2 
- 2i\kappa_2\calQ_1
+ s \, ,
\label{eq:v2aoquadrado}
\end{equation}
where $s:= -\calQ_0^{-1}(\kappa_1\calQ_2 - \calQ_1)^{2}$. 
Note that $s$ contains $\kappa_1$ alone.

Now, we insert this into (\ref{eq:iniowDf1}) and get
\begin{eqnarray}
v_1 v_4 & = & 
  -4i\kappa_1^5 \, \ren{3}\qq{0}{0}{2} 
+ 4i\kappa_1^3 \, \ren{3}\qq{0}{0}{1}
- 6\kappa_1^3\kappa_2 \, \ren{2}\q{0}{2}
\nonumber
\\
& & 
+ 2i\kappa_1^2 \kappa_3 \, \ren{2}\q{0}{0}
+ i\kappa_1\kappa_2^2 \, \ren{2}\q{0}{0}
+ 2\kappa_1\kappa_2 \,\ren{2}\q{0}{1}
\nonumber
\\
& &
+ i\kappa_1 \ren{2}\q{0}{s}
+ \kappa_1 \kappa_4 \calQ_0 \, ,
\label{eq:iniowDf2}
\end{eqnarray}
where we also used the fact that $\ren{3}\qq{0}{0}{0} =\ren{2}\q{0}{2}$.

Collecting in (\ref{eq:iniowDf2}) the terms depending only on
$\kappa_1$, we define
\begin{equation} \label{eq:defQ15}
 \calA_1 \; := \; 
  -4\kappa_1^4 \, \ren{3}\qq{0}{0}{2} 
+ 4\kappa_1^2 \, \ren{3}\qq{0}{0}{1}
+ \ren{2}\q{0}{s}
\end{equation}
and rewrite (\ref{eq:iniowDf2}) as
$$
v_1 v_4 =   
- 6\kappa_1^3\kappa_2 \, \ren{2}\q{0}{2}
+ i\kappa_1(2\kappa_1 \kappa_3 + \kappa_2^2) \, \calQ_2 
+ 2\kappa_1\kappa_2 \,\ren{2}\q{0}{1}
+ \kappa_1 \kappa_4 \calQ_0 
+i\kappa_1\calA_1 ,
$$
which is expression (\ref{eq:v1v4aux2}), as desired.
From definition (\ref{eq:defQ15}), it is evident that $\calA_1$ is
quasi-periodic, since it is a sum of quasi-periodic functions.


\subsection{Obtaining Relation (\ref{eq:v2v3aux})}
\label{ap:v2v3}

From relations (\ref{eq:v2ren})--(\ref{eq:vnren})
\begin{equation}
v_2v_3 \; = \; 
\calQ_0^{-1} \left(i\kappa_1^2\calQ_2 -i \calQ_1 +
            \kappa_2  \calQ_0 \right)\,
             \left(2i \ren{2}\q{0}{v_1v_2}  +  \kappa_3 \calQ_0
             \right) \, .
\label{eq:Diosvb}
\end{equation}
Now, by (\ref{eq:bvsiupi}), we have 
$\calQ_0^{-1}\calQ_i \ren{2}\q{0}{v_1v_2} =
\ren{2}\q{i}{v_1v_2}$. Hence, (\ref{eq:Diosvb}) becomes
\begin{eqnarray}
v_2v_3 & = & 
-2\kappa_1^2 \ren{2}\q{2}{v_1v_2}  
+2\ren{2}\q{1}{v_1v_2}  
+2i\kappa_2 \ren{2}\q{0}{v_1v_2}  
\nonumber \\
& &
+ \kappa_3\left(i\kappa_1^2\calQ_2 -i \calQ_1 +
            \kappa_2  \calQ_0 \right)\, .
\label{eq:SinueiiU}
\end{eqnarray}

Inserting (\ref{eq:2v1v2}) into the r.h.s. of (\ref{eq:SinueiiU}), it becomes
\begin{eqnarray}
v_2v_3 & = & 
-2i\kappa_1^5 \ren{2}\q{2}{2}  
+2i\kappa_1^3 \ren{2}\q{2}{1}  
-2\kappa_1^3\kappa_2 \ren{2}\q{2}{0}  
+2i\kappa_1^3\ren{2}\q{1}{2}  
\nonumber \\
& &
-2i\kappa_1\ren{2}\q{1}{1}  
+2\kappa_1\kappa_2\ren{2}\q{1}{0}  
-2\kappa_1^3\kappa_2 \ren{2}\q{0}{2}  
\nonumber \\
& &
+2\kappa_1 \kappa_2 \ren{2}\q{0}{1}
+i(2\kappa_1\kappa_2^2 + \kappa_1^2\kappa_3)\calQ_2 
-i \kappa_3\calQ_1 
+\kappa_2 \kappa_3 \calQ_0 \, .
\nonumber
\end{eqnarray}
This is (\ref{eq:v2v3aux}), as desired.


\subsection{Obtaining Relation (\ref{eq:v1v5aux})}
\label{ap:2v1v5}

According to equations (\ref{eq:v1ren})--(\ref{eq:vnren}), we have
\begin{equation} 
2v_1v_5 \; = \; 
 4i\kappa_1 \ren{2}\q{0}{v_1v_{4}+v_2v_{3}} 
+  
2\kappa_1\kappa_5 \calQ_0 \,.
\label{eq:oinuvS}
\end{equation}
By (\ref{eq:v1v4aux2}) and (\ref{eq:v2v3aux})
\begin{eqnarray}
v_1 v_4+v_2v_3 & = & 
-2i\kappa_1^5 \ren{2}\q{2}{2}  
+2i\kappa_1^3 \ren{2}\q{2}{1}  
-2\kappa_1^3\kappa_2 \ren{2}\q{2}{0}  
\nonumber \\
& &
+2i\kappa_1^3\ren{2}\q{1}{2}  
-2i\kappa_1\ren{2}\q{1}{1}  
+2\kappa_1\kappa_2\ren{2}\q{1}{0}  
\nonumber \\
& &
-8\kappa_1^3\kappa_2 \ren{2}\q{0}{2}  
+4\kappa_1 \kappa_2 \ren{2}\q{0}{1}  
+ 3i(\kappa_1^2\kappa_3 +\kappa_1\kappa_2^2 ) \calQ_2 
\nonumber \\
& & 
-i \kappa_3\calQ_1 
+( \kappa_2 \kappa_3 + \kappa_1 \kappa_4)\calQ_0 
+i\kappa_1\calA_1 \, .
\nonumber
\end{eqnarray}
Inserting this into (\ref{eq:oinuvS}), gives
\begin{eqnarray}
2v_1v_5 & = & 
8\kappa_1^6  \ren{3}\qq{0}{2}{2}  
-8i\kappa_1^4 \ren{3}\qq{0}{2}{1}  
-8i\kappa_1^4\kappa_2 \ren{3}\qq{0}{2}{0}  
\nonumber \\
& &
-8\kappa_1^4  \ren{3}\qq{0}{1}{2}  
+8\kappa_1^2  \ren{3}\qq{0}{1}{1}  
+8i\kappa_1^2\kappa_2 \ren{3}\qq{0}{1}{0}  
\nonumber \\
& &
-32i\kappa_1^4\kappa_2  \ren{3}\qq{0}{0}{2}  
+16i\kappa_1^2 \kappa_2 \ren{3}\qq{0}{0}{1}  
\nonumber \\
& & 
-12(\kappa_1^3\kappa_3 +\kappa_1^2\kappa_2^2 ) \ren{2}\q{0}{2}
+ 4\kappa_1\kappa_3\ren{2}\q{0}{1}
\nonumber \\
& & 
+4i\kappa_1( \kappa_2 \kappa_3 + \kappa_1 \kappa_4) \ren{2}\q{0}{0}
-4\kappa_1^2\ren{2}\q{0}{\calA_1 }
+  
2\kappa_1\kappa_5 \calQ_0 \,.
\nonumber
\end{eqnarray}
Let us now isolate all terms depending only on $\kappa_1$ 
and/or $\kappa_2$
in one
single term, which we call $\calA_2 $:
\begin{eqnarray}
\calA_2 & := &
8\kappa_1^6  \ren{3}\qq{0}{2}{2}  
-8i\kappa_1^4 \ren{3}\qq{0}{2}{1}  
-8i\kappa_1^4\kappa_2 \ren{3}\qq{0}{2}{0}  
\nonumber \\ & &
-8\kappa_1^4  \ren{3}\qq{0}{1}{2}  
+8\kappa_1^2  \ren{3}\qq{0}{1}{1}  
+8i\kappa_1^2\kappa_2 \ren{3}\qq{0}{1}{0}  
\nonumber \\ & &
+16i\kappa_1^2 \kappa_2 \ren{3}\qq{0}{0}{1}  
-32i\kappa_1^4\kappa_2  \ren{3}\qq{0}{0}{2}  
\nonumber \\ & &
-12 \kappa_1^2\kappa_2^2  \ren{2}\q{0}{2}
-4\kappa_1^2\ren{2}\q{0}{\calA_1 }
\label{eq:defQ22}
\end{eqnarray}
and write
\begin{eqnarray}
2v_1v_5 & = & 
\calA_2
-12\kappa_1^3\kappa_3  \ren{2}\q{0}{2}
+ 4\kappa_1\kappa_3\ren{2}\q{0}{1}
\nonumber \\ & &
+\,4i\kappa_1( \kappa_2 \kappa_3 + \kappa_1 \kappa_4) \ren{2}\q{0}{0}
+2\kappa_1\kappa_5 \calQ_0 \,.
\nonumber
\end{eqnarray}
This is (\ref{eq:v1v5aux}), as desired. Note that $\calA_2$ is clearly
quasi-periodic, being a sum of quasi-periodic functions.


\subsection{Obtaining Relation (\ref{eq:2v2v4aux})}
\label{ap:2v2v4}

Using relations (\ref{eq:v2ren})--(\ref{eq:vnren}), and
(\ref{eq:bvsiupi}) (which implies $\calQ_0^{-1}\calQ_i \ren{2}\q{0}{a}
= \ren{2}\q{i}{a}$) we have
\begin{eqnarray}
2 v_2 v_4 &=&
2\calQ_0^{-1}
\left(
             i \kappa_1^2 \calQ_2 
           - i \calQ_1 
           +  \kappa_2 \calQ_0 
\right)
\left[
2i \ren{2}\q{0}{v_1v_3}
+ i \ren{2}\q{0}{v_2^2}
+ \kappa_4 \calQ_0 
\right]
\nonumber \\
 & = &
-4 \kappa_1^2  \ren{2}\q{2}{v_1v_3}
+4             \ren{2}\q{1}{v_1v_3}
+4i\kappa_2    \ren{2}\q{0}{v_1v_3}
\nonumber \\
& &
-2  \kappa_1^2 \ren{2}\q{2}{v_2^2}
+2             \ren{2}\q{1}{v_2^2}
+2 i\kappa_2   \ren{2}\q{0}{v_2^2}
\nonumber \\
& &
+ 2\kappa_4
\left(
             i \kappa_1^2 \calQ_2 
           - i \calQ_1 
           +  \kappa_2 \calQ_0 
\right)
\nonumber
\end{eqnarray}
Inserting above the expression for $v_1v_3$ given in (\ref{eq:v1v3util}), we get
\begin{eqnarray}
2 v_2 v_4 &=&
 8  \kappa_1^6           \ren{3}\qq{2}{0}{2}
-8  \kappa_1^4           \ren{3}\qq{2}{0}{1}
-8i \kappa_1^4 \kappa_2  \ren{3}\qq{2}{0}{0}
\nonumber \\
 &  &
-4  \kappa_1^3 \kappa_3  \ren{2}\q{2}{0}
-8 \kappa_1^4           \ren{3}\qq{1}{0}{2}
+8 \kappa_1^2           \ren{3}\qq{1}{0}{1}
\nonumber \\
 &  &
+8i \kappa_1^2\kappa_2  \ren{3}\qq{1}{0}{0}
+4 \kappa_1   \kappa_3  \ren{2}\q{1}{0}
-8i \kappa_1^4 \kappa_2           \ren{3}\qq{0}{0}{2}
\nonumber \\
& &
+8i \kappa_1^2 \kappa_2           \ren{3}\qq{0}{0}{1}
-8  \kappa_1^2 \kappa_2^2         \ren{3}\qq{0}{0}{0}
+4i \kappa_1   \kappa_2 \kappa_3  \ren{2}\q{0}{0}
\nonumber \\
& &
-2  \kappa_1^2 \ren{2}\q{2}{v_2^2}
+2             \ren{2}\q{1}{v_2^2}
+2 i\kappa_2   \ren{2}\q{0}{v_2^2}
\nonumber \\
& &
+ 2\kappa_4
\left(
             i \kappa_1^2 \calQ_2 
           - i \calQ_1 
           +  \kappa_2 \calQ_0 
\right) \, .
\nonumber
\end{eqnarray}

We now isolate all terms depending only on $\kappa_1$ and/or
$\kappa_2$ in one single term, which we call $\calA_3 $:
\begin{eqnarray}
\calA_3 &:=&
 8  \kappa_1^6           \ren{3}\qq{2}{0}{2}
-8  \kappa_1^4           \ren{3}\qq{2}{0}{1}
-8i \kappa_1^4 \kappa_2  \ren{3}\qq{2}{0}{0}
\nonumber \\
 &  &
-8 \kappa_1^4           \ren{3}\qq{1}{0}{2}
+8 \kappa_1^2           \ren{3}\qq{1}{0}{1}
+8i \kappa_1^2\kappa_2  \ren{3}\qq{1}{0}{0}
\nonumber \\
 &  &
-8i \kappa_1^4 \kappa_2           \ren{3}\qq{0}{0}{2}
+8i \kappa_1^2 \kappa_2           \ren{3}\qq{0}{0}{1}
-8  \kappa_1^2 \kappa_2^2         \ren{3}\qq{0}{0}{0}
\nonumber \\
 &  &
-2  \kappa_1^2 \ren{2}\q{2}{v_2^2}
+2             \ren{2}\q{1}{v_2^2}
+2 i\kappa_2   \ren{2}\q{0}{v_2^2}
\label{eq:defQ26}
\end{eqnarray}
(note here that, by (\ref{eq:v2aoquadrado}), $v_2^2$ depends only on
$\kappa_1$ and $\kappa_2$). We can now write
\begin{eqnarray}
2 v_2 v_4 &=&
\calA_3
-4  \kappa_1^3 \kappa_3  \ren{2}\q{2}{0}
+4 \kappa_1   \kappa_3  \ren{2}\q{1}{0}
\nonumber \\
& &
+ 2i( \kappa_1^2 \kappa_4 +2 \kappa_1   \kappa_2 \kappa_3 ) \calQ_2 
  - 2i\kappa_4\calQ_1 
  +  2\kappa_2\kappa_4 \calQ_0 
 \, .
\nonumber
\end{eqnarray}
This is the desired relation (\ref{eq:2v2v4aux}). Again, note that 
$\calA_3 $ is quasi-periodic, for it is a sum of quasi-periodic functions.


\subsection{Obtaining Relation (\ref{eq:somando1})}
\label{ap:somando1}

Let us explicitly write $\kappa_1l_{n-1}$ in term {\rm (i)} of
(\ref{eq:analisechave}) using equation (\ref{eq:lntab}). We have,
$$
\kappa_1 l_{n-1}(t) \; = \; 
i \kappa_1\sum_{p=1}^{n-2}\ren{2}\q{0}{v_pv_{n-1-p}}_t
\; =: \; A(t) + B(t) \, ,
$$
where
\begin{equation} \label{eq:Ct}
A(t) \defi 2 i \, \kappa_1 
\ren{2}\q{0}{v_1v_{n-2}},
\quad
B(t) \defi i\kappa_1 \, 
\sum_{p=2}^{n-3}\ren{2}\q{0}{v_pv_{n-1-p}}
\, .
\end{equation}
The above expressions for $A(t)$ and $B(t)$ will now be worked
individually. Let us start with $A(t)$. By the inductive hypothesis, we
are allowed to use
(\ref{eq:v1ren}) and (\ref{eq:vnren}). We write
\begin{eqnarray}
A(t) &=& 2 \, i \, \kappa_1^2
\ren{2}\q{0}{i\sum_{p=1}^{n-3}\ren{2}\q{0}{v_pv_{n-2-p}} + \kappa_{n-2} \calQ_0 }
\nonumber \\
 & = &
 -2  \,\kappa_1^2
\sum_{p=1}^{n-3} \ren{3}\qq{0}{0}{v_pv_{n-2-p}} 
+
 2 \, i \,\kappa_1^2 \kappa_{n-2} 
\ren{2}\q{0}{0} \, .
\nonumber
\end{eqnarray}

Note that $A(t)$ is implicitly dependent on $\kappa_{n-3}$,
namely through $v_{n-3}$. To make this dependence explicit we have to
split the sum containing $v_{n-3}$ and write $v_{n-3}$ with the use of
(\ref{eq:vnren}):
\begin{eqnarray}
A(t) &=& 
 -2  \,\kappa_1^2
\sum_{p=2}^{n-4} \ren{3}\qq{0}{0}{v_pv_{n-2-p}} 
 -4  \,\kappa_1^2
\ren{3}\qq{0}{0}{v_1v_{n-3}} 
\nonumber \\ 
& &
+
 2 \, i \,\kappa_1^2 \kappa_{n-2} 
\ren{2}\q{0}{0}
\nonumber \\
 & = &
-2  \,\kappa_1^2
\sum_{p=2}^{n-4} \ren{3}\qq{0}{0}{v_pv_{n-2-p}} 
 -4i  \,\kappa_1^3
\sum_{p=1}^{n-4}\ren{4}\qqq{0}{0}{0}{v_pv_{n-3-p}}
\nonumber \\
 &  &
 -4  \,\kappa_1^3 \kappa_{n-3}
\ren{3}\qq{0}{0}{0}
+ 2 \, i \,\kappa_1^2 \kappa_{n-2} 
\ren{2}\q{0}{0} \, .
\label{eq:Afinal}
\end{eqnarray}

We will now work on $B(t)$, equation (\ref{eq:Ct}). 
Using (\ref{eq:v1ren}) and (\ref{eq:vnren}), we have
\begin{eqnarray} 
B(t) &=& 
2i\kappa_1 \,\ren{2}\q{0}{v_2v_{n-3}}
+i\kappa_1 \, 
\sum_{p=3}^{n-4}\ren{2}\q{0}{v_pv_{n-1-p}} \, .
\label{eq:Dtmedio}
\end{eqnarray}
Next, we have to compute separately $v_2v_{n-3}$. Using 
once more (\ref{eq:v2ren}), (\ref{eq:vnren}) 
and
(\ref{eq:bvsiupi}) (which implies $\calQ_0^{-1}\calQ_i \ren{2}\q{0}{a}
= \ren{2}\q{i}{a}$), we get
\begin{eqnarray}
v_2v_{n-3} \!\!\!&=&\!\!\! 
\calQ_0^{-1}\left(i\kappa_1^2\calQ_2 -i \calQ_1 +
            \kappa_2  \calQ_0 \right)
\left\{
i\sum_{p=1}^{n-4}\ren{2}\q{0}{v_pv_{n-3-p}}_t 
+ \kappa_{n-3} \calQ_0(t) 
\right\}
\nonumber \\
& = &
-\kappa_1^2\sum_{p=1}^{n-4}\ren{2}\q{2}{v_pv_{n-3-p}} 
+\sum_{p=1}^{n-4}\ren{2}\q{1}{v_pv_{n-3-p}}
\nonumber \\
& & 
+i\kappa_2\sum_{p=1}^{n-4}\ren{2}\q{0}{v_pv_{n-3-p}} 
+ \kappa_{n-3} \left(i\kappa_1^2\calQ_2 -i \calQ_1 +
            \kappa_2  \calQ_0 \right) \, .
\nonumber
\end{eqnarray}
This expression for  $v_2v_{n-3}$ has to be introduced into 
the first term of (\ref{eq:Dtmedio}). The result is
\begin{eqnarray}
B(t) &=& 
2i\kappa_1\sum_{p=1}^{n-4}
\Biggl\{
-\kappa_1^2\ren{3}\qq{0}{2}{v_pv_{n-3-p}} 
+\ren{3}\qq{0}{1}{v_pv_{n-3-p}} 
\nonumber \\
& &
+i\kappa_2\ren{3}\qq{0}{0}{v_pv_{n-3-p}} 
\Biggr\} +i\kappa_1 \, 
\sum_{p=3}^{n-4}\ren{2}\q{0}{v_pv_{n-1-p}} \, .
\nonumber \\
&  &
+ 2i\kappa_1\kappa_{n-3} \left[i\kappa_1^2\ren{2}\q{0}{2} -i \ren{2}\q{0}{1} +
            \kappa_2 \ren{2}\q{0}{0} \right] \, .
\label{eq:Bfinal}
\end{eqnarray}

Since both $A$ and $B$ are quasi-periodic,
we can now compute $\kappa_1M( l_{n-1})=M(A)+M(B)$. Using
(\ref{eq:Afinal}), (\ref{eq:Bfinal}), (\ref{eq:M1}) and the already
proven fact that $M\q{0}{0} = 0$, we get
\begin{equation} \label{eq:somando1b}
\kappa_1M( l_{n-1})  \; = \;  2\kappa_1 \kappa_{n-3} M\q{0}{1} +
\calR^{(1)}_n \, ,
\end{equation}
where 
\begin{eqnarray} 
\calR^{(1)}_n &\!\!\! := \!\!\!&
2i\kappa_1\sum_{p=1}^{n-4}
\Biggl\{
 -2 \kappa_1^2
M(\ren{4}\qqq{0}{0}{0}{v_pv_{n-3-p}})
+ M(\ren{3}\qq{0}{1}{v_pv_{n-3-p}})
\nonumber \\
& &
-\kappa_1^2 M(\ren{3}\qq{0}{2}{v_pv_{n-3-p}}) 
+i\kappa_2 M(\ren{3}\qq{0}{0}{v_pv_{n-3-p}}) 
\Biggr\}
\nonumber \\
& &
-2  \,\kappa_1^2
\sum_{p=2}^{n-4} M(\ren{3}\qq{0}{0}{v_pv_{n-2-p}}) 
+i\kappa_1 \, 
\sum_{p=3}^{n-4}M(\ren{2}\q{0}{v_pv_{n-1-p}}) \, .
\nonumber \\
\label{eq:R1def}
\end{eqnarray}
This is the desired relation (\ref{eq:somando1}). By inspection, one
verifies that $\calR^{(1)}_n$ depends on the constants $\kappa_1, \,
\ldots, \, \kappa_{n-4}$ only. Notice that the constant $\kappa_{n-2}$
disappeared completely when we took the mean value of $A(t) + B(t)$,
due to the crucial fact that $M(\calQ_2) = 0$. This is very important,
otherwise we would have in (\ref{eq:somando1b}) an equation for
\underline{two} unknowns $\kappa_{n-3}$ and $\kappa_{n-2}$.

\subsection{Obtaining Relation (\ref{eq:somando2})}
\label{ap:somando2}

The main point is to make the $\kappa_{n-3}$ dependence of
$l_{n-2}(t)$ explicit. Using (\ref{eq:lntab2}) and (\ref{eq:vnren})
for $v_{n-3}(t)$, we can write
\begin{eqnarray} 
q^{-1} v_2 l_{n-2} &=&
i q^{-1} v_2 \sum_{p=1}^{n-3}\ren{2}\q{0}{v_pv_{n-2-p}}
\; = \;
i  \sum_{p=1}^{n-3}\ren{2}\q{q v_2}{v_pv_{n-2-p}}
\nonumber \\
& = &
2i  \ren{2}\q{q v_2}{v_1v_{n-3}}
+i  \sum_{p=2}^{n-4}\ren{2}\q{q v_2}{v_pv_{n-2-p}}
\nonumber \\
& = &
2i\kappa_1  
\ren{2}\q{q v_2}{i\sum_{p=1}^{n-4}\ren{2}\q{0}{v_pv_{n-3-p}} + \kappa_{n-3} \calQ_0}
\nonumber \\
& &
+i  \sum_{p=2}^{n-4}\ren{2}\q{q v_2}{v_pv_{n-2-p}}
\nonumber \\
& = &
-2\kappa_1  
\sum_{p=1}^{n-4} \ren{3}\qq{q v_2}{0}{v_pv_{n-3-p}}
+i  \sum_{p=2}^{n-4}\ren{2}\q{q v_2}{v_pv_{n-2-p}}
\nonumber \\
&  &
+2i\kappa_1\kappa_{n-3}  
\ren{2}\q{q v_2}{0} \,.
\nonumber
\end{eqnarray}
According to (\ref{eq:v2ren}), 
$$
\ren{2}\q{q v_2}{0}
\; =\; 
i\kappa_1^2\ren{2}\q{2}{0}
-i\ren{2}\q{1}{0}
+ \kappa_2\ren{2}\q{0}{0}
$$ 
and, hence, $M(\ren{2}\q{q v_2}{0})=-iM\q{1}{0}=iM\q{0}{1}$. Therefore,
\begin{equation*}
M(q^{-1} v_2 l_{n-2}) 
\; = \;
-2\kappa_1\kappa_{n-3} M\q{0}{1} + \calR^{(2)}_n \, ,
\nonumber
\end{equation*}
where
\begin{equation} 
\calR^{(2)}_n \; := \; 
-2\kappa_1  
\sum_{p=1}^{n-4} M(\ren{3}\qq{q v_2}{0}{v_pv_{n-3-p}})
+i  \sum_{p=2}^{n-4}M(\ren{2}\q{q v_2}{v_pv_{n-2-p}}) \, .
\label{eq:R2def}
\end{equation}
This is the desired equation (\ref{eq:somando2}).  By inspection, one
verifies that $\calR^{(2)}_n$ depends on the constants $\kappa_1, \,
\ldots, \, \kappa_{n-4}$ only.


\section{Proof of Convergence of the $\eps$ Expansion for Periodic $f$}
\label{sec:convergence}
\zerarcounters 

Here we will present the proof of Theorem \ref{teo:convergence}, i.e.,
the proof of convergence of the $\eps$ expansion of (\ref{eq:ansatz})
for periodic $f$. It follows the ideas of \cite{pp}, but technical
adaptations are necessary. For the sake of simplification we shall
consider here only the case where $F_0=M(f)=0$. The general case
$F_0\neq 0$ can be treated following the lines described in detail in
\cite{pp}.

In terms of the Fourier coefficients $ Q_{m}$ and $Q_{m}^{(2)}$,
appearing in (\ref{eq:notQ}), of the Fourier coefficients
$V^{(n)}_{m}$ of (\ref{eq:Vnm}) and of the constants $\kappa_n$, relations
(\ref{eq:v1ren})-(\ref{eq:vnren}), become
\begin{eqnarray} 
V_{m}^{(1)} \!\!& = &\!\! \kappa_1 Q_{m} , \quad V_{m}^{(2)} \; = \;
\sum_{n_1 \in \Z \atop n_1 \neq 0} \frac{ Q_{m - n_1}\left( \kappa_1^2
    Q^{(2)}_{n_1} - \overline{Q^{(2)}_{-n_1}} \right)}{n_1 \omega} \,
+ \, \kappa_2 Q_{m },
\label{PerVFourier1}
\\
V_{m}^{(n)} \!\!\!&=&\!\!\!\!\!
       \sum_{n_1, \, n_2 \in \Z \atop n_1 + n_2 \neq 0}
       \frac{Q_{m - (n_1 + n_2)} }{(n_1 + n_2)  \omega}
         \left(\sum_{p=1}^{n-1}
         V_{n_1}^{(p)}V_{n_2}^{(n-p)} \right)
       \, + \, 
       \kappa_n Q_{m},
       \; \ \mbox{for } n \geq 3.
\label{PerVFouriern}
\end{eqnarray}
Of course, due to the choices of the constants $\kappa_n$ described
before, no secular terms appear.

By (\ref{RR1}) and by an inductive argument, we will prove the
following statement: for all $p\in \N$ and all $m\in \Z$ there are
constants $K_p>0$ such that
\begin{equation}
\left| V_{m}^{(p)}\right|\; \leq \; 
K_p\frac{e^{-\chi |m|}}{\novomod{m}^2} .
\label{hipoteseindutivaVn}
\end{equation}
To show this, let us first recall the following result, proven in
\cite{pp}:
\begin{lemma}
For $ \chi > 0$ and $ m\in\Z$ define
\begin{equation}
   \calB (m) \equiv \calB(m, \, \chi) \; := \;  \sum_{n \in \Z } 
\frac{e^{-\chi(|m-n|+|n|)}}{\novomod{m-n}^2\; \novomod{n}^2  } .
\nonumber
\end{equation}
Then one has
\begin{equation}
   \calB (m) \; \leq \; B_0 \frac{e^{-\chi|m|}}{\novomod{m}^2} ,
\nonumber
\end{equation}
for some constant $ B_0 \equiv B_0(\chi) >0$ and for all $ m\in \Z$.  
$\EndofStatement\!\!$
\label{lemaauxilialsobredecaimentodasconvolucoes}
\end{lemma}

From (\ref{PerVFourier1}) and (\ref{RR1}), we have
\begin{eqnarray}
\left| V_{m}^{(1)}\right| & \leq & 
       \calQ \frac{e^{-\chi |m|}}{\novomod{m}^2} ,
\nonumber
\\
\left|V_{m}^{(2)}\right| & \leq & 
      \frac{2\calQ^2 }{\omega}\sum_{n_1 \in \Z_\ast}
       \frac{e^{-\chi( |m-n_1|+|n_1|)}}{\novomod{m-n_1}^2\novomod{n_1}^2}
       \frac{1}{|n_1|}
           \; + \; |\kappa_2| \calQ \frac{e^{-\chi |m|}}{\novomod{m}^2},
\nonumber
\end{eqnarray}
where we used $|\kappa_1|=1$.
Now
$$
\sum_{n_1 \in \Z_\ast}
       \frac{e^{-\chi( |m-n_1|+|n_1|)}}{\novomod{m-n_1}^2\novomod{n_1}^2}
       \frac{1}{|n_1|}
\; \leq \;
\sum_{n_1 \in \Z }
       \frac{e^{-\chi( |m-n_1|+|n_1|)}}{\novomod{m-n_1}^2\novomod{n_1}^2}
\; \leq \;
B_0 \frac{e^{-\chi|m|}}{\novomod{m}^2}, 
$$
where the last inequality comes from Lemma
\ref{lemaauxilialsobredecaimentodasconvolucoes}.
Hence, we can write
$$
\left| V_{m}^{(1)}\right| \; \leq \; K_1
\frac{e^{-\chi|m|}}{\novomod{m}^2} 
\qquad \mbox{ and } \qquad 
\left| V_{m}^{(2)}\right| \; \leq \; K_2
\frac{e^{-\chi|m|}}{\novomod{m}^2},
$$
for all $m\in \Z$,
by choosing
$$
K_1 \;:= \;   \calQ  
\qquad \mbox{ and } \qquad 
K_2 \; := \;  \frac{2\calQ^2 B_0}{\omega} 
+ |\kappa_2| \calQ .
$$

To proceed, let us assume (\ref{hipoteseindutivaVn}) for 
all $p=1, \; \ldots ,\; n-1$. By (\ref{PerVFouriern}) and (\ref{RR1}),
we have
\begin{eqnarray} 
\left|V_{m}^{(n)}\right| & \leq &
       \frac{\calQ}{\omega}
 \left(\sum_{n_1, \, n_2 \in \Z \atop n_1 + n_2 \neq 0}
 \frac{e^{-\chi( |m-n_1-n_2|+|n_1|+|n_2|)}}{
 \novomod{m-n_1-n_2}^2\novomod{n_1}^2\novomod{n_2}^2}
       \frac{1 }{|n_1 + n_2|}
 \right) 
\nonumber
\\ & &
        \times \left(\sum_{p=1}^{n-1}K_pK_{n-p} \right)
        \; +\; 
       |\kappa_n| \calQ \frac{e^{-\chi |m|}}{\novomod{m}^2},
       \qquad \qquad \mbox{for } n \geq 3.
\nonumber
\end{eqnarray}
Again, applying twice Lemma
\ref{lemaauxilialsobredecaimentodasconvolucoes}, 
\begin{multline}
\sum_{n_1, \, n_2 \in \Z \atop n_1 + n_2 \neq 0}
 \frac{e^{-\chi( |m-n_1-n_2|+|n_1|+|n_2|)}}{
 \novomod{m-n_1-n_2}^2\novomod{n_1}^2\novomod{n_2}^2}
       \frac{1 }{|n_1 + n_2|} \leq 
\\
\sum_{n_1, \, n_2 \in \Z }
 \frac{e^{-\chi( |m-n_1-n_2|+|n_1|+|n_2|)}}{
 \novomod{m-n_1-n_2}^2\novomod{n_1}^2\novomod{n_2}^2} 
\; \leq \;        
(B_0)^2 \frac{e^{-\chi|m|}}{\novomod{m}^2} .
\nonumber
\end{multline}
Therefore,
$$
\left| V_{m}^{(n)}\right| \; \leq \; K_n
\frac{e^{-\chi|m|}}{\novomod{m}^2} 
$$
for all $m\in \Z$, by choosing
\begin{equation} \label{eq:estKn}
K_n \; := \;
\frac{(B_0)^2\calQ}{\omega}\left(\sum_{p=1}^{n-1}K_pK_{n-p} \right)
+ \kappa_n^0\calQ ,
\end{equation}
where $\kappa_n^0$ is some suitably chosen upper bound for
$|\kappa_n|$.  We now turn our attention to $|\kappa_n|$, for which we
have to find estimates using again the inductive hypothesis
(\ref{hipoteseindutivaVn}) for all $p=1, \; \ldots ,\; n-1$.  The
constants $\kappa_1$, $\kappa_2$ and $\kappa_3$ are are fixed by
(\ref{eq:capa123}) and $\kappa_{n}$, $n\geq 4$, are given by
(\ref{eq:capan}), from which we get
\begin{equation} \label{eq:Lpiubuw}
|\kappa_{n}| \leq 
\frac{1}{4|M\q{0}{1}|} \!
\left[
        \underbrace{\sum_{p=4}^{n-1}\left|M\left(v_p v_{n+3-p}\right)\right|}_{T_1} + 
       2\underbrace{\left|\calR^{(1)}_{n+3}\right|}_{T_2} +
       2\underbrace{\left|\calR^{(2)}_{n+3}\right|}_{T_3} + 
       2\underbrace{\left|M\left(q^{-1}v_3 l_{n} \right)\right|}_{T_4}
\right] \!,
\end{equation}
for $n \geq 4$. We have to bound each of the terms $T_i$ indicated
above. Let us start with $T_1$.

{\bf Bound for $T_1$.} 
By (\ref{eq:Vnm}), $M\left(v_p v_{n+3-p}\right) = \sum_{m\in
  \Z}V_{m}^{(p)}V_{-m}^{(n+3-p)}$. Hence, by the inductive hypothesis
(\ref{hipoteseindutivaVn}), assumed for $p=1, \; \ldots , \; n-1$, we
have
\begin{equation}
T_1 \; = \; 
\sum_{p=4}^{n-1}
\left|M\left(v_p v_{n+3-p}\right)\right| \; \leq \; 
\eta_1 \sum_{p=4}^{n-1}K_pK_{n+3-p},
\label{eq:t1final}
\end{equation}
where
$\displaystyle\eta_1 := \sum_{m\in  \Z}\frac{e^{-2\chi |m|}}{\novomod{m}^2}$.

{\bf Bound for $T_2$.} Expression (\ref{eq:R1def}) involves sums of
the mean value of functions like $\ren{k}\qq{a_1}{\cdots}{a_k}$. Let
us prove a general statement about such functions. 
\begin{proposition} \label{prop:proplegal}
For $k\geq 2$, 
let $a_1, \, \ldots , \, a_k$ be periodic functions with the same
frequency $\omega$,  and such that 
$
a_l(t) = \sum_{m\in \Z}A^{(l)}(m) e^{im \omega t}
$,
where the Fourier coefficients $A^{(l)}(m) $ satisfy 
\begin{equation}
\left|A^{(l)}(m)\right| \; \leq \; \alpha_l \frac{e^{-\chi |m|}}{\novomod{m}^2},
\label{eq:estAlm}
\end{equation}
for all $m\in \Z$ and all $l=1, \, \ldots , \, k$, where $\alpha_l>0$
and
$\chi>0$. Then, there is a positive constant $\beta_k$ such that 
the Fourier coefficients $\ren{k}\qq{a_1}{\cdots}{a_k}(m)$, $m\in \Z$,
of $\ren{k}\qq{a_1}{\cdots}{a_k}_t$ are bounded by
\begin{equation} \label{eq:koiTTuytvubU}
\left|\ren{k}\qq{a_1}{\cdots}{a_k}(m)\right| \; \leq \; 
\beta_k \, \alpha_1 \cdots \alpha_k \;
\frac{e^{-\chi |m|}}{\novomod{m}^2}  \, .
\end{equation}
$\EndofStatement$
\end{proposition}

\Proof 
Let us first consider the case $k=2$.
The Fourier coefficients of $\ren{2}\q{a_1}{a_2}_t$ are given by
\begin{equation} \label{eq:koiubU}
\ren{2}\q{a_1}{a_2}(m) \; = \; \sum_{n\in \Z}A^{(1)}(m-n)\til{A}^{(2)}(n),
\end{equation}
where
$$
\til{A}^{(2)}(m) \; := \; 
\left\{
\begin{array}{ll}
\displaystyle
\frac{A^{(2)}(m)}{im\omega}, & \mbox{ for } m\neq 0,
\\
 &
\\
\displaystyle
-\frac{1}{i\omega}\sum_{k\in\Z_*}\frac{A^{(2)}(k)}{k} ,
& \mbox{ for } m=0
\end{array} \, .
\right.
$$
From (\ref{eq:estAlm}), it follows that 
$$
\left|A^{(2)}_I(m)\right| \; \leq \; 
\alpha_2\frac{D_0}{\omega} \frac{e^{-\chi |m|}}{\novomod{m}^2},
$$
where $D_0 :=\sum_{m\in\Z}\frac{e^{-\chi |m|}}{\novomod{\;m\;}^2}$.
Therefore, from (\ref{eq:koiubU}), by (\ref{eq:estAlm}) and by Lemma
\ref{lemaauxilialsobredecaimentodasconvolucoes}, one has
$$
\left|\ren{2}\q{a_1}{a_2}(m) \right|\; \leq \;
\alpha_1 \alpha_2\frac{B_0 D_0 }{\omega}
\;\frac{e^{-\chi |m|}}{\novomod{m}^2}
.
$$
This proves the statement for $k=2$. The general case follows from
(\ref{eq:ionviPOI}), by induction. \QED

As a corollary, one sees from (\ref{RR1}) that the functions 
$\calQ_0$, $\calQ_1$ and $\calQ_2$ have Fourier coefficients bounded
as $|\calQ_i (m)| \leq \gamma_i \frac{e^{-\chi |m|}}{\novomod{\;m\;}^2}$
for some positive $\gamma_i$. Moreover, by the inductive hypothesis
(\ref{hipoteseindutivaVn}) and by Lemma
\ref{lemaauxilialsobredecaimentodasconvolucoes}, the Fourier
coefficients $(v_p v_q)(m)$ of product functions like $v_p(t) v_q(t)$,
with $p, \; q =1, \, \ldots , \, n-1$, are also bounded as
\begin{equation} \label{eq:ynpUbvu}
\left|(v_p v_q)(m)\right| \; \leq \;
B_0 \, K_p\, K_q \, \frac{e^{-\chi |m|}}{\novomod{m}^2}
\end{equation}
for all $m\in \Z$. The consequence of all this is that 
for indices $i_j \in \{0, \; 1, \; 2\}$ and $p, \; q =1, \, \ldots ,
\, n-1$
one has
$$
\left|
\ren{k}\left( i_1| i_2| \cdots | i_{k-1}|v_p v_q\right)(m)
\right|
\;\; \leq \;\;
\gamma_{i_1, \, i_2, \, \cdots , \, i_{k-1}}\;  K_p\, K_q\; 
\frac{e^{-\chi |m|}}{\novomod{m}^2},
\qquad \forall m\in\Z,
$$
from some positive constants $\gamma_{i_1, \, i_2, \, \cdots , \,
  i_{k-1}}$, depending on the indices  $i_j$. 

Returning our attention to expression (\ref{eq:R1def}), we conclude
that 
\begin{equation}
\left|\calR^{(1)}_{n+3}\right| \;\; \leq \;\;
\eta_2\sum_{p=1}^{n-1}K_p K_{n-p}
+\eta_3\sum_{p=2}^{n-1} K_p K_{n+1-p}
+\eta_4\sum_{p=3}^{n-1}K_p K_{n+2-p}  \, ,
\label{eq:t2final}
\end{equation}
for certain positive constants $\eta_2, \; \eta_3, \;\eta_4$.

{\bf Bound for $T_3$.}
Since $qv_2$ is a linear combination of the functions
$\calQ_0$, $\calQ_1$ and $\calQ_2$, we conclude from (\ref{eq:R2def})
and from the previous arguments that
\begin{equation}
\left|\calR^{(2)}_{n+3} \right| \; \leq \; 
\eta_5
\sum_{p=1}^{n-1} K_pK_{n-p} 
+  
\eta_6
\sum_{p=2}^{n-1} K_pK_{n+1-p} \, ,
\label{eq:t3final}
\end{equation}
for certain positive constants $\eta_5, \; \eta_6$.

{\bf Bound for $T_4$.}
By (\ref{eq:lntab2}), one has
$$
M(q^{-1}v_3 l_{n}) \; = \; 
i \sum_{p=1}^{n-1}M(\ren{2}\q{qv_3}{v_pv_{n-p}}) .
$$
From (\ref{eq:v3especial}) and (\ref{eq:ynpUbvu}), we see that both
$qv_3$ and $v_pv_{n-p}$ satisfy the conditions of Proposition
\ref{prop:proplegal}. Hence, by (\ref{eq:koiTTuytvubU}), $
M(\ren{2}\q{qv_3}{v_pv_{n-p}}) \leq \eta_7 K_p K_{n-p}$ for some
positive constant $\eta_7$ and
\begin{equation}
M(q^{-1}v_3 l_{n}) \; \leq \;
 \eta_7\sum_{p=1}^{n-1} K_p K_{n-p}.
\label{eq:t4final}
\end{equation}

We are finished with the bounds for the terms $T_i$ of
(\ref{eq:Lpiubuw}). If we collect (\ref{eq:t1final}),
(\ref{eq:t2final}), (\ref{eq:t3final}) and (\ref{eq:t4final}) and
return to (\ref{eq:estKn}), we conclude that there are positive
constants $\Gamma_1, \; \Gamma_2, \; \Gamma_3, \; \Gamma_4$ such that
we can recursively define
\begin{multline}
K_n \; := \; 
 \Gamma_1\sum_{p=1}^{n-1} K_pK_{n-p} 
+\Gamma_2\sum_{p=2}^{n-1} K_p K_{n+1-p}
\\
+\Gamma_3\sum_{p=3}^{n-1}K_p K_{n+2-p}
+\Gamma_4\sum_{p=4}^{n-1}K_pK_{n+3-p} \, ,
\nonumber
\end{multline}
for $n> 4$, after fixing the convenient values for $K_1$, $K_2$, $K_3$
and $K_4$. Note that we can choose $K_1 = K_2 =K_3=K_4$ taking
$K_i = \max\{K_1, \; K_2, \;K_3, \; K_4\}$ for all $i=1, \,
\ldots , \, 4$.
Defining $\Gamma:=\max\{\Gamma_1, \; \Gamma_2, \; \Gamma_3, \; \Gamma_4
\}$, we can redefine the $K_n$ so as to have the more convenient choice
\begin{multline}
K_n \; := \; 
 \Gamma
\left[
\sum_{p=1}^{n-1} K_pK_{n-p} 
+\sum_{p=2}^{n-1} K_p K_{n+1-p}
\right. 
\\ 
\left.
+\sum_{p=3}^{n-1}K_p K_{n+2-p}
+\sum_{p=4}^{n-1}K_pK_{n+3-p} 
\right]
\, .
\label{eq:FIUybouF}
\end{multline}

Expression (\ref{eq:FIUybouF}) has an analogous one in \cite{pp}, with
the distinction that only the two first sums above occurred in the
corresponding expression for $K_n$. From now on, we follow closely the
steps of \cite{pp}. The first one is to show that $K_n$ is a {\it
  non-decreasing} sequence.  We have
\begin{multline}
K_{n+1} -K_n \; := \; 
\Gamma \left[
\left(\sum_{a=1}^4 K_a\right)K_{n}
+
\left(\sum_{a=1}^3 K_a \right)(K_n-K_{n-a})
\right.
\\
\left.
+\sum_{p=4}^{n-1}K_p(K_{n+4-p} -K_{n-p}) 
\right] .
\nonumber
\end{multline}
Therefore, assuming inductively
$K_1 =  K_2 = K_3 = K_4 \leq \ldots \leq K_n$ implies $K_n \leq K_{n+1}$, thus 
proving that the sequence is non-decreasing.
Next, we write (\ref{eq:FIUybouF}) as
\begin{equation*}
K_n =  
\Gamma
\left[
\sum_{a=1}^3 \left( \sum_{b=a}^3 K_b\right)K_{n-a} 
+\sum_{p=4}^{n-1}K_p\left(K_{n-p}+ K_{n+1-p}+K_{n+2-p} +K_{n+3-p} \right)
\right].
\nonumber
\end{equation*}
Since the sequence is non-decreasing, we have
$K_{n-p} + K_{n+1-p} +K_{n+2-p} + K_{n+3-p} \leq 4K_{n+3-p}$ and $K_{n-a}
\leq K_{n-1}$ for $a = 1, 2, 3$. Hence, we may say that
\begin{eqnarray}\label{eq:Ppuin}
K_n & \leq & \Gamma \sum_{a=1}^{3}\left(\sum_{b=a}^{3} K_b \right) 
K_{n-1} + 4\Gamma\sum_{p=4}^{n-1}K_p K_{n+3-p} 
\nonumber \\
& = & \til{\Gamma} \, K_{n-1} K_4 + 4\Gamma\sum_{p=4}^{n-1}K_p K_{n+3-p} \, ,
\end{eqnarray}
where ${\displaystyle \til{\Gamma} := \frac{\Gamma}{K_4} \sum_{a=1}^{3}
\left(\sum_{b=a}^{3}K_b\right)}$ is a positive constant. Adding up the
positive quantity ${\displaystyle \til{\Gamma}\sum_{p=4}^{n-2}K_p
K_{n+3-p}}$ to (\ref{eq:Ppuin}) and setting $\Lambda :=
\max{\{\til{\Gamma}, \; 4\Gamma\}}$, we get
\begin{equation}\label{eq:Ppuin2}
K_n \; \leq \; \Lambda \sum_{p=4}^{n-1}K_p K_{n+3-p} \, .
\end{equation}

Let us now define another auxiliary sequence $J_k$ such that 
$J_l=K_l$ for $l=1, \; 2, \; 3, \; 4$ and
$$
J_n \; := \; 
 \Lambda\sum_{p=4}^{n-1}J_pJ_{n+3-p}\, .
$$
for $n>4$. It is a simple exercise to show from (\ref{eq:Ppuin2}) that
$K_n \leq J_n$ for all $n$.  Now, let us consider the translated
sequence $ L_{n}=J_{n+2} $, $n\geq 1$.  We have
\begin{equation}
L_n \; = \; 
\Lambda\sum_{p=4}^{n+1}J_pJ_{n+5-p}
\; = \; 
\Lambda\sum_{p=4}^{n+1}L_{p-2}L_{n+3-p}
\; = \; 
\Lambda\sum_{p=2}^{n-1}L_{p}L_{n+1-p}\,
.
\label{eq:definicaodasequanciaLn}
\end{equation}

The sequence $ {\mathbf c}_n$ defined by $ \displaystyle {\mathbf c}_n =
\sum_{p=2}^{n-1} {\mathbf c}_p {\mathbf c}_{n-p+1} $ for $ n\geq 3$,
with $ {\mathbf c}_1 = {\mathbf c}_2 =1$, defines the so-called
``Catalan numbers'', which can be expressed in a closed form as
\begin{equation}
     {\mathbf c}_n \; = \; \frac{(2n-4)!}{(n-1)!(n-2)!} , \qquad n\geq
     2 \, .
\nonumber
\end{equation}
By Stirling's formula, the ${\mathbf c}_n$'s have the following
asymptotic behaviour: $ {\mathbf c}_n \approx
\frac{1}{16\sqrt{\pi}}\;\frac{4^n}{n^{3/2}} $, for $n$ large.  The
existence of a connection between the Catalan numbers and the sequence
$ L_n$ is evident from (\ref{eq:definicaodasequanciaLn}). Two
distinctions are the factor $\Lambda$ appearing in
(\ref{eq:definicaodasequanciaLn}) and the fact that $L_1=L_2 =
K_3=K_4$ are not necessarily equal to $ 1$.  One can, however, easily
show that $ L_n = (K_3)^{n-1}\;\Lambda^{n-2}\;{\mathbf c}_n $, $n\geq
2$.  Hence, the following asymptotic behaviour can be established:
\begin{equation}
L_n \; \approx \;
\frac{1}{16\sqrt{\pi}K_3\Lambda^2} 
\;\;\frac{(4K_3\Lambda)^n}{n^{3/2}},
\qquad n \mbox{ large.}
\nonumber
\end{equation}

Since $K_n \leq J_n = L_{n-2}$, we conclude that for $n$ large
$K_n \leq M_0 (M_1)^n$, for some positive 
constants $M_0, \; M_1$. This completes the proof of
Theorem~\ref{teo:convergence}. $\QED$


\section{Comments on the Fourier Coefficients}
\label{app:novo}
\zerarcounters

Relations (\ref{U11U12}) and (\ref{u11u12}) can be obtained in our
case by repeating the analysis of \cite{pp}. One of the conclusions is
that the Fourier coefficients of (\ref{u11u12}) are analytic functions
of $\eps$, for $|\eps|$ small enough. There is, however, a point to be
noticed here. If we follow the steps of \cite{pp} and compute the
Fourier expansion of $R(t)^{-2}$ to obtain the Fourier expansion for
$S(t)$, this last function will contain terms like $\int_0^t
e^{2i\Omega t}dt$, which behave like $\Omega^{-1}$. In case (I),
treated in \cite{pp}, this is not problematic, since $\Omega =
O(\eps)$ and since such terms are always multiplied by a factor $\eps$
(see (\ref{definicaodeU})). In our case, however, $\Omega = O(\eps^3)$
and we have to look such terms more carefully.

Looking at the $\eps$-expansion for $U_{12}(t)$, we have
$$
U_{12}(t)= -ie^{-i\gamma(\eps)}
\left[
\eps W_1(t) + \eps^2 W_2(t) + O(\eps^3)
\right],
$$
where, after a lengthy computation, we get 
$$
\eps W_1(t) \; = \;
\eps \sum_{n, \, m} 
\frac{ Q_{n}^{(2)}}{i(n\omega + 2\Omega)}
\left[
\overline{Q_{n-m}}
e^{i(m\omega + \Omega)t}
-
\overline{Q_{-m}}
e^{i(m\omega - \Omega)t}.
\right]
$$
The problematic term is that proportional to $\eps/\Omega$, which
appears for $n=0$. However, this term is proportional to
$Q_{0}^{(2)}=M(\calQ_0)=0$, by hypothesis and, hence, the Fourier
coefficients of $\eps W_1$ are analytic on $\eps$. Let us look now at
$\eps^2W_2$. After another lengthy computation, one gets
\begin{eqnarray*}
\eps^2 W_2(t) \!\!\! & = &
\!\!\!\!
\frac{2\kappa_1\eps^2}{i\omega}
\sum_{p, \, n, \, m \atop m\neq 0}
\left[
\frac{ Q_{n-m}^{(2)} \overline{Q_{n-p}} Q_{m}^{(2)} 
}{  m(n\omega +2\Omega) }
e^{i(p\omega + \Omega)t}
-
\frac{ Q_{n-m}^{(2)} \overline{Q_{-p}} Q_{m}^{(2)} 
}{  m(n\omega +2\Omega) }
e^{i(p\omega - \Omega)t}
\right]
\\
\!\!
& + &
\!\!\!\!
\frac{\kappa_1\eps^2}{i\omega}
\sum_{p, \, n, \, m \atop m\neq 0}
\left[
\frac{ Q_{n}^{(2)} \overline{Q_{m-p}} Q_{m}^{(2)} 
}{  m(n\omega +2\Omega) }
e^{i(p\omega - \Omega)t}
-
\frac{ Q_{n}^{(2)} \overline{Q_{m+n-p}} Q_{m}^{(2)} 
}{  m(n\omega +2\Omega) }
e^{i(p\omega + \Omega)t}
\right].
\end{eqnarray*}
Again, the dangerous terms are those with $n=0$, since the are 
proportional to $\eps^2/\Omega$. However, one sees by inspection that
such terms vanish in the expression above. In fact, in the last two
sums they appear proportional to $Q_{0}^{(2)}$, which is zero. 
In the first two sums, they appear proportional to 
$\sum_{m\neq 0}\frac{Q_{-m}^{(2)}Q_{m}^{(2)} }{m}$, which is also
zero, as one sees by interchanging $m\to -m$. 

Our conclusion is that all dangerous terms like $1/\Omega$ appear
multiplied at least by factors $\eps^3$ and are, therefore, analytic.

\section{An Identity on Sums of Bessel Functions}
\label{identidadeBessel}

Here we will sketch the proof of identity (\ref{eq:identidadeBessel}),
since we did not found mention to it in the literature.  Using the
well know identity (due to Schl\"afli and Gegenbauer) for products of
Bessel functions (see \cite{Watson})
\begin{equation}
J_\mu(z)J_\nu(z) =\frac{1}{\pi}\int_{-\pi/2}^{\pi/2}J_{\mu+\nu}(2z\cos\theta)
\cos((\mu-\nu)\theta)\,d\theta , 
\label{eq:idprodBessel}
\end{equation}
one gets
$$
S_m(x) \; := \; \sum_{k\neq 0}\frac{J_{m+k}(x)^2}{k}
\; = \;
(-1)^m\int_{-\pi/2}^{\pi/2}J_0(2x\cos\theta)g(\theta)e^{2i m \theta}\, d\theta,
$$
where
$
g(\theta)  :=  \frac{1}{\pi}\sum_{k\neq 0}(-1)^k \frac{\sin(2k\theta)}{k}
$. On the interval $(-\pi/2, \; \pi/2)$ one has
$g(\theta)=-2\theta/\pi$. Hence,
\begin{eqnarray*}
S_m(x) & = &
\frac{2(-1)^m}{\pi}
\int_{-\pi/2}^{\pi/2}J_0(2x\cos\theta)\theta\sin(2 m \theta)\, d\theta
\\
& = &
\frac{(-1)^{m+1}}{\pi}
\left.\left(
\frac{d}{d\nu}\int_{-\pi/2}^{\pi/2}J_0(2x\cos\theta)\cos(2 \nu \theta)\, d\theta
\right)\right|_{\nu =m} .
\end{eqnarray*}
Using again (\ref{eq:idprodBessel})
(with $\mu=-\nu $), one finds
$$
S_m(x) \; = \;
(-1)^{m+1}
\left. 
\frac{\partial}{\partial\nu} 
\left[J_{\nu}(x)J_{-\nu}(x) \right]
\right|_{\nu=m}
\; = \;
J_m(x)
\left[ 
\left. -2\frac{\partial}{\partial \nu}J_{\nu}(x)\right|_{\nu=m} +
\pi Y_m(x)
\right].
$$
This is (\ref{eq:identidadeBessel}), as desired,
where $Y_m$ is the Bessel function of second kind, given by
$\pi Y_m(x) = \left. \left( \frac{\partial}{\partial \nu}J_{\nu}(x)
    -(-1)^m\frac{\partial}{\partial \nu}J_{-\nu}(x)
  \right)\right|_{\nu=m}$, for integer $m$.

\end{appendix}


\vspace{0.7cm}

\begin{flushleft}
J. C. A. Barata and D. A. Cortez \\
Universidade de S\~ao Paulo \\
Instituto de F\'{\i}sica \\
Caixa Postal 66318 \\
S\~ao Paulo - 05315 970 - SP - Brasil
\end{flushleft}


\end{document}